\documentstyle[12pt]{article}
\newskip\humongous \humongous=0pt plus 1000pt minus 1000pt

\newif\ifdtup

\catcode`\@=11
\@addtoreset{equation}{section}
\def\theequation{\arabic{section}.\arabic{equation}}
\def\@normalsize{\@setsize\normalsize{15pt}\xiipt\@xiipt
\abovedisplayskip 14pt plus3pt minus3pt%
\belowdisplayskip \abovedisplayskip
\abovedisplayshortskip \z@ plus3pt%
\belowdisplayshortskip 7pt plus3.5pt minus0pt}
\def\small{\@setsize\small{13.6pt}\xipt\@xipt
\abovedisplayskip 13pt plus3pt minus3pt%
\belowdisplayskip \abovedisplayskip
\abovedisplayshortskip \z@ plus3pt%
\belowdisplayshortskip 7pt plus3.5pt minus0pt
\def\@listi{\parsep 4.5pt plus 2pt minus 1pt
     \itemsep \parsep
     \topsep 9pt plus 3pt minus 3pt}}
\relax
\catcode`@=12
\catcode`\@=11
\def\section{\@startsection{section}{1}{\z@}{3.5ex plus 1ex minus
   .2ex}{2.3ex plus .2ex}{\large\bf}}

\def\thesection{\arabic{section}}

\def\appendix{\setcounter{section}{0}
 \def\thesection{Appendix \Alph{section}}
 \def\theequation{\Alph{section}.\arabic{equation}}}
\newcommand{\beq}{\begin{equation}}
\newcommand{\eeq}{\end{equation}}
\newcommand{\bea}{\begin{eqnarray}}
\newcommand{\eea}{\end{eqnarray}}
\newcommand{\beas}{\begin{eqnarray*}}
\newcommand{\eeas}{\end{eqnarray*}}

\newcommand{\non}{\nonumber}
\def\de{\partial}

\def\de{\partial}
\def\om{\Omega}

\def\bede{\begin{description}}
\def\ede{\end{description}}

\begin{document}
\begin{titlepage}
\begin{center}
{\Large
Improved Convergence Proof of the Delta Expansion and Order Dependent
Mappings}
\end{center}
\vspace{1ex}
\begin{center}
{\large
Riccardo Guida$^{1,3}$, Kenichi Konishi$^{2,1,3}$ and Hiroshi Suzuki$^{4}$}
\end{center}
\vspace{1ex}
\begin{center}
{\it $^{1}$ Dipartimento di Fisica -- Universit\`a di Genova\\
     Via Dodecaneso, 33 -- 16146 Genova, Italy\\
     $^{2^*}$ LPTHE, Univ.~de Paris Sud, Centre d'Orsay\\
     B\^at.~211, 91405  Orsay, Cedex, France\\
     $^{3}$ Istituto Nazionale di Fisica Nucleare -- sez.~di Genova\\
     Via Dodecaneso, 33 -- 16146 Genova, Italy\\
     $^{4}$ Department of Physics -- Ibaraki University, Mito 310, Japan }
\end{center}
\begin{center}
e-mail: guida@ge.infn.it, konishi@ge.infn.it, hsuzuki@mito.ipc.ibaraki.ac.jp
\end{center}
\medskip
{\bf ABSTRACT:}
We improve and generalize in several accounts the recent rigorous
proof of convergence of delta expansion - order dependent mappings
(variational perturbation expansion) for the energy eigenvalues of
anharmonic oscillator. For the single-well anharmonic oscillator
the uniformity of convergence in $g\in[0,\infty]$ is proven. The
convergence proof is extended also to complex values of $g$ lying
on a wide domain of the Riemann surface of $E(g)$.
Via the scaling relation \`a la Symanzik, this proves the convergence
of delta expansion for the double well in the strong coupling regime
(where the standard perturbation series is non Borel summable), as well
as for the complex ``energy eigenvalues'' in certain metastable
potentials. Sufficient conditions for the convergence of delta expansion
are summarized in the form of three theorems, which should apply to a
wide class of quantum mechanical and higher dimensional field theoretic
systems.

\vfill
\begin{flushleft}
GEF-Th-4/1995; LPTHE Orsay-95/25; IU-MSTP/3;
hep-th/9505084 \hfill May 1995
\end{flushleft}
\begin{flushleft}
$^*$ Laboratoire associ\'e au C.N.R.S. URA-D0063
\end{flushleft}
\smallskip
\end{titlepage}
\section{Introduction}
\label{sec:intro}
A recent rigorous proof of the convergence of scaled delta expansion
\cite{GKS}, (inspired by and developing further an earlier work by
Duncan and Jones \cite{DJ}), has provided for the first time a solid basis
for understanding the success of the delta expansion - order dependent
mappings (DE-ODM in the following).

However the proof given there has several unsatisfactory features.
First, the proof does not apply as it stands to the double well case
($\omega^2<0$) while numerical evidence suggests that (optimized) delta
expansion converges for coupling constant above some critical value
(i.e., the central barrier lower than a critical height).

Also, delta expansion converges numerically well even in the
$g\to\infty$ limit \cite{DJ,wjanke}, but the proof given in
Ref.~\cite{GKS} is valid for finite $g$ only.
(In other words, uniformity of the convergence in the whole $g$ ($>0$)
axis was not proven). Even for finite $g$ the absolute value of the
remainder was numerically found to be smaller than our upper bound,
suggesting that a tighter bound was possible to get, even if the bound
found there was sufficient for proving the convergence for any finite $g$.

The proof of \cite{GKS} furthermore does not apply if the coupling
constant $g$ is taken to be complex. As the energy eigenvalues of the
anharmonic oscillator and those of the double well are related by
analytic continuation in $g$, it is interesting to know the domain in the
complex $g$ plane in which delta expansion converges. Seznec and
Zinn-Justin \cite{SZ} analyzed these problems and argued in a non
rigorous fashion that delta expansion or order dependent mappings
converges in some complex domain of~$g$.

These shortcomings of our previous proof have clearly to do with the
fact that the detailed knowledge on the analytic property of the energy
eigenvalue as a function of complex $g$, as explored by Bender and Wu
\cite{BW},
Loeffel and Martin \cite{LM}, Simon \cite{Simon}, has not been fully
exploited in the proof.\footnote{We thank J. Zinn-Justin for stressing
this point to us.}

In the present paper, we improve on these points and generalize our
convergence proof in several accounts. First we generalize the allowed
form of the reparametrization (conformal transformation) of the coupling
constant used to construct DE-ODM. Secondly, the DE-ODM for the
anharmonic oscillator/double well system is proved to converge for
complex coupling constant $g$ in a domain on the Riemann surface
of $E(g)$, similar to the one suggested by Seznec and Zinn-Justin
\cite{SZ}. Also the proof is considerably simplified with respect to
our previous work.

Furthermore, a general set of sufficient conditions for the convergence
of DE-ODM will be given for a generic physical quantity considered as
a function of the coupling constant and whose standard perturbative
series is known.

\smallskip
To set the general background for the present work, some known facts
about the scaling relation \`a la Symanzik, the relation between the
energy eigenvalues in the single well and those in the double well
oscillator, and the general analytic structure of $E(\omega^2,g)$ in
$g$, will be briefly reviewed below.

Consider the anharmonic oscillator with $q^{2M}$ potential described by
the Hamiltonian,
\beq
   H={p^2\over2}+{\omega^2\over2}q^2+{g\over2M}q^{2M}.
\label{aho}
\eeq
By considering the scale transformation $q\to\xi q$, unitarily
implemented on the Hilbert space of states (as first suggested by Symanzik),
one finds \cite{LM,Simon} a remarkable relation for the energy,
considered as a function of $\omega$ and $g$,
\beq
   E(\omega^2,g)=\xi^{-2}E(\xi^4\omega^2,\xi^{2M+2}g)
\label{scaling1}
\eeq
valid in the analyticity region for any complex number $\xi$ ($\ne0$).
This relation plays a powerful role in the study of analytic properties
of $E(\omega^2,g)$.

An example of Eq.~(\ref{scaling1}) is
\beq
   E_K(\omega^2,g)=g^{1/(M+1)}E_K(g^{-2/(M+1)}\omega^2,1)
\label{scaling2}
\eeq
valid for each level $K$. Combining this with the knowledge that the
expansion in $\omega^2$ yields an analytic regular perturbation series,
one proves \cite{Simon} the (uniform) convergence of the large $g$ expansion,
\beq
   E_K(1,g)/g^{1/(M+1)}=\sum_{n=0}^\infty d_n(g^{-2/(M+1)})^n,
\label{largeg}
\eeq
for {\it complex\/} $g$ such that
\beq
   |g|\ge g_0>0
\label{strongconvreg}
\eeq
(note that $g_0$ can always be increased so as to ensure the convergence
at the border).

Also, the choice $\xi=e^{\pi i/2}$ leads to a periodicity relation for
energy,
\beq
   E(\omega^2,e^{(M+1)\pi i}g)=-E(\omega^2,g).
\label{period}
\eeq

Furthermore, by an appropriate choice of $\xi$ in Eq.~(\ref{scaling1})
it is possible to relate the energy eigenvalues in the single well
potential to those in the double well potential. Indeed, the choice
$\xi=e^{i\pi/4}$ leads to
\bea
   E^{\rm(DW)}_K (\omega^2,g)&\equiv&E_K(-\omega^2,g)
\non\\
   &=&e^{-i\pi/2}E_K(\omega^2,e^{i(M+1)\pi/2}g).
\label{scaling3}
\eea

Analytic continuation of $E(g)\equiv E(1,g)$ and the structure of its
singularities in complex $g$ plane have been studied with different
techniques and levels of rigor: Bender and Wu \cite{BW} and Bender {\it et
al.} \cite{Bender} used in their pioneering works the WKB approximation,
together with some general arguments; Loeffel and Martin \cite{LM}
studied analytic continuation of the exact solutions of differential
Schr\"odinger equation, and Simon \cite{Simon} used the general theory
of linear operators in Hilbert spaces.

Some of their main results are:
\bede
\item{i)}
$E(g)$ is analytic in the whole cut $g$ plane \cite{LM,Simon};
\item{ii)}
the Riemann surface has a global $M+1$-sheeted structure:
a full rotation of angle $2(M+1)\pi$ at fixed $|g|$ brings back to
the original value of $E(g)$ \cite{BW,Simon};
\item{iii)}
$E(g)$ has an infinite number of pairs of isolated square-root
type branch points\footnote{\label{rigorousfoot}
A rigorous proof of the detailed structure of higher Riemann sheets is still
lacking (to our knowledge): Simon \cite{Simon} proved that the only isolated
singularities of $E(g)$ are algebraic branch points with no negative powers,
but he could only assume the non existence of non-isolated singularities other
than $g=0$ and of other pathologies.}
(Bender-Wu singularities) \cite{BW} on the symmetric positions with
respect to the phase $\pm i(M+1)\pi/2$ that accumulate towards $g=0$
(in some Riemann sheets other than the first), with asymptotic phase
$e^{\pm i(M+1)\pi/2}$ \cite{BW,Simon}; at these singularities in complex
$g$ plane the crossing of energy levels occurs.
\item{iv)}
A full closed tour of angle $2(M+1)\pi$ in $g$ plane starting from $E_K(g)$
but this time by crossing the Bender-Wu cuts several times, takes one
back to the energy eigenvalue of a different level, $E_{K+n}(g)$, where
$n$ depends on the number and the way the Bender-Wu cuts have been
crossed \cite{BW}.
\item{v)}
For small enough $g$, $E(g)$ is analytic in a sectorial domain
\cite{Simon}: let $0<\theta<(M+1)\pi/2$ then there is a $G$ such that
$E(g)$ is analytic in the subset
\begin{equation}
   \left\{\,g\mid0<|g|<G,\,|{\rm Arg}\,g|<\theta\,\right\}
\end{equation}
of the Riemann surface. Also in any such sector the Rayleigh-Schr\"odinger
perturbation theory is asymptotic uniformly in the angle \cite{Simon}.
\item{vi)}
A unique strong coupling expansion Eq.~(\ref{largeg}) exists which
converges uniformly for $|g|$ large \cite{Simon}.
\ede

The rest of the paper is organized as follows. We first recall in
Sec.~\ref{sec:deodm} the relation between the delta expansion
and the order dependent mappings, which serves also to review briefly
these ideas and at the same time to define our convention.
The convergence proofs are given in Sec.~\ref{sec:proof}, which
constitutes the main result of this paper.
In Sec.~\ref{sec:opti} convergence of the {\it optimized\/} expansion
is discussed. The case of general $q^{2M}$ oscillators is discussed
in Sec.~\ref{sec:q2M}.
Sec.~\ref{sec:green} discusses briefly the convergence of delta
expansion for the Green's functions of the anharmonic oscillator.
Summary and Discussion is given in Sec.~\ref{sec:summary}.
Several technical issues and examples are collected in Appendices:
\ref{sec:properties} discusses some properties of the conformal
transformation used in the proof, \ref{sec:zero} the convergence proof
of DE-ODM in a zero-dimensional analogue model. The condition on the
trial parameter following from the optimization condition is discussed
in \ref{sec:optimization}. In \ref{sec:appd} we present the calculation of
critical exponents in $\phi_3^4$ theory by use of DE-ODM. Finally a
reconstruction of the strong coupling expansion coefficients from the
standard perturbation series is discussed in \ref{sec:appe}.

\section{Delta expansion versus order dependent mapping}
\label{sec:deodm}
The delta expansion is a method to perform a systematic calculation,
taking as the zeroth approximation an unperturbed action
$A_0(\{\Omega\})$ which contains a set of trial ``vacuum'' parameters
$\{\Omega\}$ (e.g., trial frequency, mass).\footnote{
In this respect the delta expansion is close in spirit to the Hartree
or self consistent approximation. There are many works related to delta
expansion. Standard earlier references can be found in \cite{GKS,DJ}.
See \cite{recent} for more recent ones.}
Perturbation with respect to the rest of the action
$\delta[A-A_0(\{\Omega\})]$
is calculated in the standard fashion, and the $N$th order result
(with $\delta=1$) $S_N^{\rm(DE)}$ is computed by fixing $\{\Omega\}$
{\it order by order\/} by some appropriate criterion, e.g., by an
optimization procedure.

For the quartic quantum mechanical oscillator Eq.~(\ref{aho}) with $M=2$,
the ``unperturbed'' and ``interaction'' Hamiltonians can be taken as
\beq
   H_0(\om)
   ={p^2\over2}+{\om^2\over2}q^2;\quad
   H_I(\om)={\omega^2-\om^2\over2}q^2+{g\over4}q^4.
\eeq
In this case $S_N^{\rm(DE)}$ for the energy eigenvalues was proved
\cite{GKS} to converge to the exact answer as $N\to\infty$ if the trial
frequency is scaled with the order $N$ as
\beq
   x_N\equiv{\Omega_N\over\omega}=CN^\gamma,
\label{scaling}
\eeq
where either
\beq
   1/3<\gamma<1/2,\quad C>0,
\label{index}
\eeq
or
\begin{equation}
   \gamma=1/3,\quad C\geq\alpha_c g^{1/3},\quad
   \alpha_c=0.5708751028937741\cdots.
\label{newindex}
\end{equation}
Also the divergence of delta expansion with $x_N$ lying outside the
above range, was explicitly shown. The convergence of the ``optimized''
delta expansion (in which $\Omega_N$ is determined for instance by
the condition $\partial S_N/\partial\Omega_N=0$, at each order),
is a consequence of the above (see Sec.~\ref{sec:opti} below).

Formally, the delta expansion in this case can be generated by a substitution,
\bea
   &&\omega\to
   \Omega\left(1+\delta\,{\omega^2-\Omega^2\over\Omega^2}\right)^{1/2}
   =\omega x[1-\delta\beta(x)]^{1/2},
\nonumber\\
   &&g\to\delta\cdot g,
\non\\
   &&x\equiv{\Omega\over\omega},\quad\beta(x)\equiv 1-{1\over x^2},
\label{sub}
\eea
in the standard perturbation series
\bea
   E^{\rm(pert)}(\omega^2,g)
   &=&\omega\sum_{n=0}^{\infty}c_n\left({g\over\omega^3}\right)^n,
\non\\
   &=&\omega E^{\rm (pert)}(1,g/\omega^3),
\label{per}
\eea
followed by the expansion in $\delta$ (and $\delta=1$; $\omega=1$ at the
end).

In the order dependent mapping method of Seznec and Zinn-Justin \cite{SZ},
one considers a physical quantity $S(g)$ regarded as the function of a
(in general complex) coupling constant $g$, and whose perturbative
series in powers of $g$ is known up to the $N$th order.
One makes then a conformal transformation,
\beq
   g=\rho F(\beta)
\eeq
to the new (in general complex) variable $\beta$.
The (order by order-) adjustable parameter of the transformation, $\rho$, is
kept always real and positive. The transformation is such that $g=0$ and
$g=\infty$ are mapped to $\beta=0$ and $\beta=1$ respectively;
the full range $[0,\infty]$ in $g$ is mapped to $[0,1]$ in the space of
$\beta$.
For instance, in the case of the anharmonic oscillator the choice
\beq
   F(\beta)\equiv{\beta\over(1-\beta)^\alpha},
\label{conformal}
\eeq
with $\alpha>0$ a fixed parameter, will work. The $N$th order
approximation $S_N^{\rm(ODM)}$ is obtained by re-expanding $S(g)$
in powers of $\beta$ up to $N$th order,
\beq
   S(g)\to S_N^{\rm(ODM)}=\sum_{n=0}^NP_n(\rho)\beta^n,
\eeq
and $\rho$ is fixed by some criterion, order by order. In Ref.~\cite{SZ}
Seznec and Zinn-Justin adopt the criterion of the fastest apparent
convergence, i.e., they require $\rho$ at order $N$ to be a root of
$P_{N+1}(\rho)=0$.

In the case of the energy eigenvalues of the quartic anharmonic
oscillator, the same authors considered the function $\Psi(\beta,\rho)$
related to the energy by
\beq
   E(1,g)={1\over(1-\beta)^{1/2}}
   \left[(1-\beta)^{1/2}E(1,g(\beta))\right]
   \equiv{1\over(1-\beta)^{1/2}}\Psi(\beta,\rho);
\label{ODM}
\eeq
with a particular conformal transformation
\beq
   g(\beta)=\rho{\beta\over(1-\beta)^{3/2}};
   \quad\beta\equiv1-{1\over x^2};\quad\rho={g\over x^3\beta(x)}.
\label{ctras}
\eeq
Only the function $\Psi(\beta,\rho)$ (but not the prefactor
$1/(1-\beta)^{1/2}=x$) is re-expanded in $\beta$.

Comparison between the set of equations (\ref{sub}) and (\ref{per}),
and the set Eqs.~(\ref{ODM}) and (\ref{ctras}), shows that in this
case the delta expansion and order dependent mapping (with the particular
conformal transformation chosen above) coincide formally.
$\beta$ in the two cases can be identified.

It is clear that the two approaches are closely related to each other.
The original form of the delta expansion has a certain intuitive appeal,
being physically motivated. However it is not obvious how to apply such
a method to higher dimensional field theory models where mass
renormalization is important. The order dependent mappings, on the other
hand, being mainly mathematically motivated, appear to be more flexible.
In particular the conformal transformation in the latter can be of
a more general form than suggested by the shift of the quadratic term.
In the present paper these methods will be referred to indistinguishably as
DE-ODM.

\section{Convergence proof}
\label{sec:proof}
In this section convergence proofs of DE-ODM are given. The convergence
in the case of the quartic potential ($M=2$) for finite positive $g$
is given in Sec.~\ref{sec:proof1}, for a wide class of conformal
transformations, Eqs.~(\ref{conf1})--(\ref{alpha}).
In Sec.~\ref{sec:proof2}, the convergence proof for quartic potential
is generalized to complex values of $g$ (including higher Riemann sheets
of $E(g)$) by use of a particular conformal transformation
($\alpha =3/2$ in Eq.~(\ref{conf1})). All the (sufficient) conditions
used for the convergence proof of DE-ODM are summarized in
Sec.~\ref{sec:proof3} in the form of three theorems to be applied for
a generic function $S(g)$ of~$g$.

The strategy for the proof of convergence is common in all cases,
and can be summarized as follows (with reference to the energy of anharmonic
oscillator).
\bede
\item{1)}
Consider the energy as a function of the new variable $\beta$,
$E(g(\beta))$, through the conformal transformation,
\beq
   g=\rho F(\beta)=\rho{\beta\over(1-\beta)^\alpha};\quad
   (\rho{\rm\ real\ positive}).
\label{conf1}
\eeq
For consistency with delta expansion notation (see Sec.~\ref{sec:deodm})
it will be useful to define also the variable $x$ by:
\beq
   \beta\equiv 1-{1\over x^2};\quad
   \rho={g\over x^{2\alpha}\beta}.
\label{conf2}
\eeq
The parameter $\alpha$ will be always chosen in the range
\beq
   1<\alpha\leq2
\label{alpha}
\eeq
to ensure some properties essential for the proof (see \ref{sec:properties}).
Since $E(g)$ is analytic in the (open) cut $g$ plane
(see Sec.~\ref{sec:intro}),
the composed function $E(g(\beta))$ is analytic (at least) in its (open) image
$\cal D$ in $\beta$ plane.
It is shown in \hbox{\ref{sec:properties}} that $\cal D$ is the interior
of an apple like figure minus the negative real axis.
(See Fig.~1 for the case $\alpha=3/2$.)
\item{2)}
Introduce a function $\Psi(\beta,\rho)$ by
\beq
   E(g(\beta))=
   {1\over(1-\beta)^\sigma}\Psi(\beta,\rho).
\label{Psi}
\eeq
$E(g)$ in the case of the energy of $q^{2M}$-oscillator is known to
behave as $g^{1/(M+1)}$ at $g\to\infty$ (see Sec.~\ref{sec:intro}); the
extraction of the factor $1/(1-\beta)^\sigma$ is made so that the
function $\Psi(\beta,\rho)$ is bounded at $\beta=1$. The
condition for that is:
\beq
   \sigma\ge{\alpha\over M+1}.
\label{sigma}
\eeq
In cases we are interested in the convergence proof for values of $g$
in higher Riemann sheets, more special choices of $\sigma$ will be
required; see below.

By construction $\Psi(\beta,\rho)$ is analytic in $\cal D$.
\item{3)}
Write Cauchy's theorem in the $\beta$ plane (See Fig.~1),
\begin{equation}
   \Psi(\beta)
   =\oint_\Gamma{d\beta'\over2\pi i}{\Psi(\beta')\over\beta'-\beta},
\label{cauchy}
\end{equation}
where $\Gamma$ is a closed contour which traces the boundary of the
domain $\cal D$ (we assume that the energy is continuous on the two
edges of the $g$ cut, see below).
Note also that the contribution to the Cauchy integral from the portion
of $\Gamma$ that turns around $\beta=0$ is negligible because the
uniform asymptoticity of the perturbative series for $E(g)$ implies
that $E(g)\to1/2$ for $|g|\to0$, see Sec.~\ref{sec:intro}). In certain
cases one can (and must) enlarge further the contour in the $\beta$
plane (depending on the analytic structure of $\Psi(\beta,\rho)$);
such a deformation is crucial in the proof of convergence in higher sheets.
\item {4)}
Taylor expand $\Psi(\beta,\rho)$ in $\beta$ up to the $N$th
order\footnote{Derivatives respect to $\beta$ in $\beta=0$ are taken
{}from the positive axis and can be exchanged with the integral in
$\beta'$ of Cauchy formula essentially due to regularity of $\Psi$
in this region (in particular the behaviour of ${\rm Disc}\,\Psi$ for
$\beta'\to0^-$, see Sec.~\ref{sec:proof1}).}
\begin{eqnarray}
   \Psi_N(\beta,\rho)&=&\sum_{n=0}^N{1\over n!}
   \partial^n_\beta\Psi(0,\rho)\beta^n
\non\\
&=&
   \sum_{n=0}^N\oint_\Gamma{d\beta'\over2\pi i}
   {\Psi(\beta')\over\beta'}\left({\beta\over\beta'}\right)^n.
\label{taylor}
\end{eqnarray}
The remainder for the energy, $R_N(\beta)=[\Psi(\beta,\rho)-\Psi_N(\beta,
\rho)]/(1-\beta)^\sigma$ is then given by
\begin{equation}
   R_N(\beta)={\beta^{N+1}\over(1-\beta)^\sigma}
   \oint_\Gamma{d\beta'\over2\pi i}
   {\Psi(\beta',\rho)\over(\beta')^{N+1}(\beta'-\beta)}.
\label{remainder}
\end{equation}
Study of various contributions (essentially from the two regions
$\beta'\sim1$ and $\beta'\sim0$) in Eq.~(\ref{remainder}) in the limit
$N\to\infty$, $\rho=g/(x^{2\alpha}\beta)\to0$, leads to convergence
proofs and the related conditions for various parameters.
\ede

\subsection{Proof for the quartic oscillator ($M=2$) for real $g$}
\label{sec:proof1}
We first prove the convergence of DE-ODM to the energy eigenvalues of
the quartic oscillator ($M=2$ in Eq.~(\ref{aho})) for any finite
positive value of $g$. Following the general steps outlined above, we
make the conformal transformation
\beq
   g=\rho F(\beta)=\rho{\beta\over(1-\beta)^\alpha};\quad
   (\rho{\rm\ real\ positive}).
\label{confo1}
\eeq
with $1<\alpha\le2$. Finite positive real $g$ will be mapped in
$\beta\in(0,1)$ by this conformal transformation.

We introduce the function $\Psi(\beta,\rho)$ by $\Psi(\beta,\rho)
=(1-\beta)^\sigma E(g(\beta))$ (with $\sigma$ restricted as in
Eq.~(\ref{sigma})) and use the Cauchy's theorem with the contour given
in Fig.~1.

To analyze the remainder
\begin{equation}
   R_N(\beta)={\beta^{N+1}\over(1-\beta)^\sigma}
   \oint_\Gamma{d\beta'\over2\pi i}
   {\Psi(\beta')\over(\beta')^{N+1}(\beta'-\beta)},
\end{equation}
we divide the contour on $\beta$ plane in two parts: $\Gamma_A$ the part
which wraps around the subset $[\beta_c\equiv-1/(\alpha-1),0]$ of the
negative real axis, and $\Gamma_B=\Gamma/\Gamma_A$ the rest of the
contour, the upper and lower arms of the apple like curves in Fig.~1
(a complete study of $\Gamma$ for generic $\alpha$ is given in
\ref{sec:properties}).

Consider first the contribution from $\Gamma_A$. Clearly the factor
$(1-\beta')^\sigma$ in
\beq
   \Psi(\beta',\rho)=(1-\beta')^\sigma E(g(\beta'))
\eeq
takes the same value on the upper and lower sides of the cut along
$\beta\in[\beta_c,0]$. Hence the discontinuity (imaginary part) of
$\Psi$ can be expressed in terms of that of $E$ and this part of the
remainder can be written as
\beq
   R_N^{(A)}=
   {\beta^{N+1}\over(1-\beta)^\sigma}
   \int_{\beta_c}^0{d\beta'\over\pi}
   {(1-\beta')^\sigma\,{\rm Im}\,E(g(\beta'))
   \over(\beta')^{N+1}(\beta'-\beta)}.
\eeq
In the limit $N\to\infty$, $\rho\to0$, we have
\beq
   g(\beta')=\rho{\beta'\over(1-\beta')^\alpha}\to0^-
\eeq
for any value of $\beta'\in\Gamma_A$. Thus the known $g\to0^-$ behavior
of ${\rm Im}\,E(g)$ \cite{BW},
\beq
   {\rm Im}\,E(g)\sim C_E(-g)^{-c}\exp\left[-{a\over(-g)^{1/b}}\right]
\eeq
can be used in the integrand ($b\equiv M-1=1$, $a=4/3$ here and $c$ is
a positive constant depending on the energy level considered). One then
finds
\beq
\label{integral}
   |R_N^{(A)}|\le
   {\rm const.}\,{|\beta|^{N}\over|1-\beta|^\sigma}\rho^{-c}
   \int_0^{|\beta_c|}d\beta'{\beta'}^{-(N+1+c)}
   \exp\left[-{a\over \rho\beta'}\right],
\eeq
where the constant (${\rm const.}$) is independent of the order
of DE-ODM and of $g$,\footnote{
Any such constant will be indicated this way in the following.}
and inequalities
\beq
   |\beta-\beta'|\ge|\beta|
\label{strictg}
\eeq
and
\bea
   &&|1-\beta_c|\ge|1-\beta'|\ge1,
\non\\
   &&\rho{|\beta'|\over|1-\beta_c|^\alpha}\le|g(\beta')|\le\rho|\beta'|
\label{gbound}
\eea
valid for $\beta'\in\Gamma_A$, $\beta\in[0,1]$ have been used.
Taking an upper bound by raising the upper limit of integration to
infinity one gets:
\bea
   |R_N^{(A)}|&\le&
   {\rm const.}\,{|\beta|^{N}\over|1-\beta|^\sigma}\rho^{-c}
   \left({\rho\over a}\right)^{N+c}\Gamma(N+c)
\\
   &\sim&(\rho N)^N
    \sim\left({N\over x^{2\alpha}}\right)^N,
\label{boundA}
\eea
where use was made of the asymptotic relation,
\beq
   \beta=1-{1\over x^2}\sim1-\left({\rho\over g}\right)^{1/\alpha}.
\label{asymptotic}
\eeq

Next consider the contribution from $\Gamma_B$. The range of $\alpha$,
$1<\alpha\le2$, has been chosen so that the following inequalities
\bea
   &&|\beta-\beta'|\ge|1-\beta|;
\label{betabound1}
\\
   &&|\beta'|\ge1,
\label{betabound2}
\eea
hold for $\beta'$ on $\Gamma_B$ (see \ref{sec:properties})
and $\beta\in[0,1]$.

Furthermore we {\it assume\/} the continuity of $E(g)$ on the two edges
of the negative $g$ cut. This physically reasonable assumption is
motivated by the fact that all the known singularities of $E(g)$ in the
nearby domain
\bea
   &&g=|g|e^{i\theta};\qquad|g|>0
\non\\
   &&\pi\le\theta\le{3\pi\over 2},\quad
   {\rm or}\quad-{3\pi\over2}\le\theta\le-\pi
\eea
are simple square-root type branch points \cite{BW} (see footnote
\ref{rigorousfoot}). Also, due to the fact that the energy is analytic
on the cut plane, that $E(g)\sim g^{1/3}$ for large $g$, and that $E(g)$
is bounded in the limit $g\to0$, the following bound holds on the cut plane
including the two edges of the negative $g$ axis:
\beq
   |E(g)|<(A+B|g|^{1/3}).
\eeq
In terms of $\Psi$ we have (on the whole of $\Gamma\cup{\cal D}$):
\beq
   |\Psi(\beta,\rho)|
   <\left[A+B
   \left({\rho|\beta|\over|1-\beta|^\alpha}\right)^{1/3}\right]
   |1-\beta|^\sigma
\eeq
{}from which (by use of Eq.~(\ref{sigma})) one easily derives
\beq
   |\Psi(\beta',\rho)|<{\rm const.};\quad\beta'\in\Gamma_B.
\label{psibound}
\eeq

It follows from Eqs.~(\ref{betabound1}), (\ref{betabound2}) and
(\ref{psibound}) that the contribution from $\Gamma_B$ to $R_N$,
$R_N^{(B)}$, is bounded by
\beq
   |R_N^{(B)}|\le{\rm const.}\,{|\beta|^{(N+1)}\over|1-\beta|^{1+\sigma}}
   \sim e^{-N/x^2}
   \sim\exp\left[-N\left({\rho\over g}\right)^{1/\alpha}\right]
\label{boundB}
\eeq
(where Eq.~(\ref{asymptotic}) has been used).

It clear from Eqs.~(\ref{boundA}) and (\ref{boundB}) that both pieces
of the remainder, $R_N^{(A)}$ and $R_N^{(B)}$, vanish in the limit of
large $N$ if $x$ or $\rho$ scale with $N$ as
\beq
   x=CN^\gamma;\quad{1\over2\alpha}<\gamma<{1\over2}\quad(C>0)
\label{gamma}
\eeq
or
\beq
  \rho={C'\over N^{\gamma'}};\quad1<\gamma'<\alpha\quad(C'>0).
\label{mu}
\eeq
Accordingly DE-ODM for an energy eigenvalue for the quartic oscillator,
constructed with the class of conformal transformations in Eq.~(\ref{confo1})
with
\beq
   1<\alpha\leq2;
\eeq
and
\beq
   \sigma\ge{\alpha\over3};
\label{sigma2}
\eeq
converges pointwise to the exact answer for any positive $g$,
if $x$ (or $\rho$) is scaled as above.

Finally, for completeness we state here that the convergence also holds
if $x$ (or $\rho$) is scaled precisely as
\beq
   x=CN^{1/(2\alpha)};\quad{\rm or}\quad\rho={C'\over N};
\label{extremex}
\eeq
with
\beq
   C\ge\left({3u_*\over4}\right)^{1/(2\alpha)}g^{1/(2\alpha)};
   \quad{\rm or}\quad C'\le{4\over3u_*}.
\label{extremeC}
\eeq
This is a special case of the Theorem~3 treated in Sec.~\ref{sec:proof3}
and $u_*$ is a solution of the set of equations (\ref{phieq1}) and
(\ref{phieq2}) (with $b=1$). In this case convergence is ensured
by the factor $|\beta|^N$ appearing in both pieces of the remainder.
For a particular conformal transformation with $\alpha=3/2$, it reduces
to the condition with $(3u_*/4)^{1/3}=0.5708751089\cdots$, found earlier
\cite{GKS}.

\smallskip
\noindent{\bf Remarks}
\bede
\item{(i)}
The result of this subsection generalizes considerably the result found
earlier (the case with $\alpha=3/2$) by the present authors \cite{GKS}
and argued for by others \cite{SZ}, allowing a class of conformal
transformations to be used in the construction of DE-ODM. The
convergence proof is also simpler than our original one.
\item{(ii)}
As is suggested from the bounds Eq.~(\ref{boundA}) and Eq.~(\ref{boundB}),
the convergence proof of this section can be actually extended to any complex
$g$ in the first Riemann sheet satisfying ${\rm Re}\,1/g^{1/\alpha}>0$.
See Theorem~1 in Sec.~\ref{sec:proof3}.
\ede

\subsection{Proof for the anharmonic oscillator/double well system
($M=2$) with complex $g$}
\label{sec:proof2}
We consider the quartic oscillator ($M=2$) again but now the coupling
constant $g$ is allowed to lie anywhere in the Riemann surface of $E(g)$.
Through the scaling relation Eqs.~(\ref{scaling1}) and
(\ref{scaling3}) this permits us to study simultaneously both
the single and double well problems.

As before we make the conformal transformation,
\beq
   g=\rho F(\beta)=\rho{\beta\over(1-\beta)^\alpha};
   \quad(\rho{\rm\ real\ positive}).
\label{conf1'}
\eeq
and introduce the function $\Psi$ by
\beq
   E(g(\beta))={1\over(1-\beta)^\sigma}\Psi(\beta,\rho).
\label{Psi2}
\eeq

The key observation which permits us to extend the proof to $g$ lying on
higher Riemann sheets, is the following. The existence of the large $g$
expansion for $E(g)$,
\beq
   E_K(1,g)=g^{1/3}\sum_{n=0}^\infty d_n(g^{-2/3})^n,
\label{largeg'}
\eeq
with a convergence radius $g_0$, shows that for the following particular
choice of parameters
\beq
   \sigma={\alpha\over3};
   \quad\alpha={3\over2}k,\,(k=1,2,3,\cdots)
\label{ancond}
\eeq
the function $\Psi(\beta,\rho)$ is analytic\footnote{\label{weierstrass}
By use of Weierstrass' theorem on series of uniformly convergent analytic
functions.} in $\beta$ inside a certain little circle $\cal C$ of radius
$r_0$ centered at $\beta=1$, as well as in the domain $\cal D$. The
first of the conditions in Eq.~(\ref{ancond}) undoes the branch point
at $g=\infty$ due to the overall $g^{1/(M+1)}$ factor; the second
eliminates the same branch point due to the fractional powers of $g$
in the expansion parameter~\cite{SZ}.

Note that such elimination of the branch point at $\beta=1$ is possible
only for a particular choice of the conformal transformation. Indeed,
the condition for analyticity in $\cal C$, Eq.~(\ref{ancond}),
together with the condition on the exponent $\alpha$ of the
transformation, $1<\alpha<2$ (which is needed to keep $\beta'$ along
$\Gamma_B$ outside the unit circle), implies that the only possible
value for $k$ in Eq.~(\ref{ancond}) is $k=1$. The only allowed conformal
transformation is the one with
\begin{equation}
   \alpha={3\over 2}.
\end{equation}
{}From the first of Eq.~(\ref{ancond}) it also follows that $\sigma=1/2$.
These particular values are assumed hereafter in this subsection.

The radius $r_0$ of $\cal C$ is easily found to be, in the limit
$N\to\infty$,
\beq
   r_0\sim\left({\rho\over g_0}\right)^{2/3},
\label{r0}
\eeq
by use of Eq.~(\ref{asymptotic}).

Since now the function $\Psi(\beta,\rho)$ is analytic in the joint
region ${\cal D}\cup{\cal C}$, the contour $\Gamma$ can be enlarged
to $\Gamma'$ so as to trace its boundary. See Fig.~2. As will be seen
shortly, this apparently minor deformation of the integration contour
is crucial. A further enlargement of the radius of $\cal C$
is {\it not possible\/} because the images of the Bender-Wu branch
points appear just outside $\cal C$.\footnote{
$g_0$ is larger than the largest absolute value of the Bender-Wu branch
points. For otherwise there would be values of $g$ such that $|g|>g_0$
and for which the strong coupling expansion for $E/g^{1/3}$ does not
converge, which is a contradiction.}
The smallest distance from one of those to the point $\beta=1$ shrinks
to zero together with $r_0$ as $N\to\infty$.

Before proceeding further, some subtleties involved in analytic
structure of the function $\Psi$ and in the consequent enlargement
of the contour $\Gamma\to\Gamma'$ should be explained. First let us note
that conformal transformation $g=\rho\beta/(1-\beta)^{3/2}$ introduces
the doubling of the number of the global Riemann sheets (six) in
$E(g(\beta))$ considered as a function of $\beta$ as compared to
the situation for $E(g)$ (three). As is clear from the form of the
strong coupling expansion Eq.~(\ref{largeg'}), there are a square-root
branch point at $\beta=1$ and a (global) cubic-root branch point at
$\beta=0$ ($\beta=\infty$ is a sixth root type branch point).

A detailed study of the conformal mapping from the $g$ Riemann surface
to $\beta$ Riemann surface shows the following correspondence.
Two values of $\beta$ with the same (geometrical) position on a pair of
$\beta$ Riemann sheets, connected by going around $\beta=1$ once
(which may be called ``mirror'' Riemann sheets), are mapped to values of
$g$ differing by the phase $e^{3\pi i}$. For example, the domain $\cal D$
of Fig.~2 corresponding to the first and its mirror (second) sheet, is the
image of the cut plane $-\pi<{\rm Arg}\,g <\pi$ and the angular region
$2\pi<{\rm Arg}\,g<4\pi$, respectively. Analogously, the part of the
circle $\cal C$ which lies outside $\cal D$ (Fig.~2), represents the
image of the region $\pi<{\rm Arg}\,g < 2\pi$; $|g|>g_0$ and that of
$4\pi<{\rm Arg}\,g <5\pi=-\pi$; $|g|>g_0$.

Now, the analyticity of $\Psi$ in $\cal C$ around $\beta=1$ implies, due to
the uniqueness of analytic continuation, that the function $\Psi$ takes
exactly the same value at corresponding points in each pair of mirror
Riemann sheets.

In terms of $E=(1-\beta)^{1/2}\Psi$ this implies a relation
\beq
   E(e^{3\pi i}g)=-E(g)
\label{qui}
\eeq
for any $g$. (This relation of course follows immediately from the
scaling relation, see Eq.~(\ref{period}). Nonetheless the above
derivation, which essentially hinges upon the existence of a strong
coupling expansion, can be used in a more general situation:
See Sec.~\ref{sec:proof3}.)

The true meaning of ``enlarging the contour from $\Gamma$ to $\Gamma'$''
can be appreciated now: As $\beta'$ moves along $\Gamma'$ {\it twice},
first in the first sheet and continuing in the second sheet
and then back, its image $g'$ goes through the full $g$ Riemann surface
once, as shown in Fig.~3.

Nevertheless, in applying Cauchy's theorem for $\Psi(\beta,\rho)$ one
may ignore the double looping of the $\Gamma'$ contour: on whichever sheet
a given $\beta$ may lie, the integration over one of the loops vanishes,
the factor $1/(\beta'-\beta)$ being analytic inside it.

\smallskip
Coming back to the convergence proof, for any $\beta$ lying inside
${\cal D}\cup{\cal C}$ we study the remainder
\begin{equation}
   R_N(\beta)={\beta^{N+1}\over(1-\beta)^{1/2}}
   \oint_{\Gamma'}{d\beta'\over2\pi i}
   {\Psi(\beta')\over(\beta')^{N+1}(\beta'-\beta)}.
\end{equation}
By splitting again the contour in two parts, $\Gamma_A$ the part
wrapping around the negative real axis (common to $\Gamma$ and
$\Gamma'$), and $\Gamma'_B=\Gamma'/\Gamma_A$, the rest of the contour
(see Fig.~2).

The contribution from $\Gamma_A$ can be treated as done in
Sec.~\ref{sec:proof1}, except that $\alpha=3/2$ and variables $g$ and
$\beta$ are complex now. By use of Eq.~(\ref{gbound}) and of
\beq
   |\beta-\beta'|\ge K_\beta
   \equiv{\rm max}\,(|{\rm Im}\,\beta|,{\rm Re}\,\beta),
\eeq
one finds easily
\bea
   |R_N^{(A)}|&\le&
   {\rm const.}\,{|\beta|^{N+1}\over K_\beta |1-\beta|^{1/2}}\rho^{-c}
   \left({\rho\over a}\right)^{N+c}\Gamma((N+c))
\non\\
   &\sim&{(\rho N)^N \over|1-\beta|^{1/2}}
   \sim|g|^{1/3}(\rho N)^N,
\label{newboundA}
\eea
where an asymptotic relation $\beta\sim1-(\rho/g)^{2/3}$ has been used.

The appearance of $K_\beta$ in the denominator of the first line of
Eq.(\ref{newboundA}) means that the values of $\beta$ in
${\cal D}\cup{\cal C}$ which are arbitrarily close to the real
negative axis, must be excluded. We shall see shortly that the remainder
vanishes as $N\to\infty$ only for $g$ lying on a certain region of the
Riemann surface (Eq.~(\ref{complexg})). For any $g$ there and for
any $N\ge N_0$ (with $N_0$ independent of $g$) the inequality
$K_\beta\ge{\rm const.}\,|\beta|$ holds, hence the convergence
will indeed be uniform in $g$ in that region.

There is an important difference in the estimate of $R_N^{(B)}$
due to the enlargement of the contour from $\Gamma$ to $\Gamma'$.
For $\beta'$ on $\Gamma'_B$, the following inequality holds:
\beq
   |\beta'|\ge1+r_0\cos(\phi_\alpha)+O(r_0^2)
\label{newbetabound}
\eeq
where
$\phi_\alpha\equiv\pi(1-1/\alpha)$ is the angle between the apple
like curve and the positive $\beta$ axis in $\beta=1$, see
\ref{sec:properties}. For the generic range $1<\alpha<2$ of $\alpha$,
the angle $\phi_\alpha$ lies in the region $0<\phi_\alpha<\pi/2$
so that $\cos\phi_\alpha>0$. For the particular conformal transformation
with $\alpha=3/2$ we are considering here, $\phi_\alpha=\pi/3$;
hence $\cos\phi_\alpha=1/2$.

Also, to bound the factor $1/|\beta'-\beta|$, one must restrict $g$ to
range in a complex domain such that the corresponding variable $\beta$
satisfies
\beq
   |\beta'-\beta|\ge\epsilon r_0\qquad\forall\beta'\in\Gamma'_B,
\label{newbetabound2}
\eeq
where $\epsilon$ is a small positive fixed ($\beta$ independent)
constant of order $O(1)$. Appearance of $r_0$ (radius of $\cal C$)
in the condition above is due to the fact that for any fixed finite $g$
its image $\beta$ approaches $1$ in the limit $N\to\infty$ and that
the nearest $\beta'$ are those on the boundary of $\cal C$.
In terms of $g$, it is sufficient that $g$ is restricted to be separated
{}from the inverse image of the contour $\Gamma'_B$ (Fig.~3), by a fixed
($g$ independent) finite distance. It will be seen below that the
condition for the (uniform) exponential vanishing of the remainder
automatically ensures such a condition to be satisfied.

Furthermore, since $E(g)\sim g^{1/3}$ at large $g$ on the whole Riemann
surface, $\Psi$ is bounded in the neighborhood of $\beta=1$
(see Eq.~(\ref{sigma})):
\beq
   |\Psi(\beta',\rho)|<{\rm const.};\quad\beta'\in\Gamma'_B.
\label{psibound2}
\eeq
Collecting these results one gets
\bea
   |R_N^{(B)}|&\le&{\rm const.}{|\beta|^{N+1}\over|1-\beta|^{1/2}r_0}
   \left({1\over 1+r_0/2}\right)^{N+1}
\non\\
   &\sim&|g|^{1/3}
   \exp\left[-N\rho^{2/3}\left({\rm Re}\,{1\over g^{2/3}}
                               +{1\over2g_0^{2/3}}\right)\right],
\label{newboundB}
\eea
where Eq.~(\ref{r0}) has been used. Note the presence of the
$g$-independent exponential factor in Eq.~(\ref{newboundB}) which
was absent in Eq.~(\ref{boundB}).

{}From Eqs.~(\ref{newboundA}) and (\ref{newboundB}) one sees that the full
remainder $R_N^{(A)}+R_N^{(B)}$ vanishes (uniformly)
in the limit $N\to\infty$ with
\beq
   \rho={C'\over N^{\gamma'}}\quad{\rm with}\quad1<{\gamma'}<3/2,
\eeq
if the complex values of $g$ are restricted in (any closed subset of)
the region
\beq
   {\rm Re}\,{1\over g^{2/3}}+{1\over 2 g_0^{2/3}}>0.
\label{complexg}
\eeq

The regions {\it excluded\/} by such a condition are one heart-like
closed domain ${\cal H}_1$
\beq
   |g|^{2/3}\le-2g_0^{2/3}\cos{2\theta\over3};\quad
   {3\pi\over4}<\theta<{9\pi\over4};\quad
   \theta\equiv{\rm Arg}\,g,
\eeq
and an analogous domain ${\cal H}_2$ with $\theta\to-\theta$. See
Fig.~3. The boundary curves of these domains cross the straight lines
corresponding to ${\rm Arg}\,g =\pi$, $2\pi$, $4\pi$ ($=-2\pi$), $5\pi$
($=-\pi$) at modulus $|g|=g_0$.

As is clear from Fig.~3, the (inverse) image of the contour $\Gamma'$
lies inside ${\cal H}_1\cup{\cal H}_2$. It follows that the condition
for the uniform vanishing of the remainder, Eq.~(\ref{complexg}),
automatically guarantees Eq.~(\ref{newbetabound2}).

We have thus proved that DE-ODM for an energy eigenvalue of the quartic
oscillator, constructed with the conformal mapping with $\alpha=3/2$ and
with $\sigma=1/2$, uniformly converges to the correct analytic
continuation $E(g)$ in every closed subset of the region
\beq
    {\rm Re}\,{1\over g^{2/3}}+{1\over2g_0^{2/3}}>0,
\label{complexg'}
\eeq
if $\rho$ or $x$ is scaled as
\beq
  \rho={C'\over N^{\gamma'}};\quad1<{\gamma'}<{3\over 2},
\label{mu'}
\eeq
or
\beq
   x=cg^{1/3}N^\gamma;\quad{1\over 3}<\gamma<{1\over 2}
\label{gamma'}
\eeq
($C'$ or $c$ is real and independent of $g$ and $N$).

As in the previous case, the convergence holds for the extreme case of
scaling with
\bea
   &&\rho={C'\over N};\quad{\rm or}\quad x=cg^{1/3}N^{1/3},
\non \\
   &&c>c_*=0.5708751089\cdots\quad{\rm or}\quad C'<{1\over c_*^3}.
\label{extreme}
\eea

\smallskip
\noindent{\bf Remarks}
\bede
\item{(i)}
The $g$ independent factor in the exponent of Eq.~(\ref{newboundB})
shows that even for real finite $g$, the bound for the remainder is
actually numerically smaller than suggested by Eq.~(\ref{boundB}),
although the exponential behavior
($|R|<\exp(-{\rm const.}\,N^{1-2\gamma'/3})$) is the same. The convergence
is fastest for the choice $\gamma'=1$.
\item{(ii)}
The same $g$ independent factor is especially important for the strong
coupling limit, $g\to\infty$. By using Eqs.~(\ref{newboundA}) and
(\ref{newboundB}) we prove that DE-ODM converges to the correct answer for
$\lim_{g\to\infty}E(g)/g^{1/3}$, if $x$ or $\rho$ is scaled according to
Eq.~(\ref{gamma'}) or to Eq.~(\ref{extreme}), and if the limit
$g\to\infty$ is taken at each order in $N$, {\it before\/} $N$ is sent
to $\infty$. This remarkable fact was noted earlier empirically.
\item{(iii)}
As is noted in Introduction, $E(g)$ continued analytically to
$g=|g|e^{3\pi i/2}$ is (apart from a phase factor $-i$) equal to the
energy eigenvalue of a double well. The result above proves\footnote{
The $N$th order delta expansion approximant for the double well is
precisely \cite{GKS} given by the substitution $x^2\to-x^2$
(that is, $\omega^2\to-\omega^2$) made in the corresponding approximant
of the single-well anharmonic oscillator. It follows that the remainder
for the double well is correctly given by $R_N^{(A)}+R_N^{(B)}$ apart
{}from an overall factor $i$.}
that DE-ODM for the quantum double well converges to the correct answer
as long as $|g|>2^{3/2}g_0$. This explains the empirical fact that lower
order (optimized) delta expansion for the double well gives good
approximate results for coupling constant above some critical value
(or the central barrier below a critical height) but fails below
it.\footnote{
This was noted independently by Okopinska \cite{Oko}, by Kleinert
\cite{KleiDW}, by the present authors (unpublished), and by
E.~Hofstetter (private communication).}
{}From our own numerical study of DE-ODM for the double well up to
$N\simeq50$, we find a numerical value $g_0\simeq0.10--0.15$.
Due to the fact that Eq.~(\ref{complexg'}) refers to an {\it upper bound\/}
of the remainder, this should be considered as an order of magnitude
estimate for the convergence radius of the strong coupling expansion.
See the following remark.
\item{iv)}
Our result depends on the knowledge of analyticity of $E(g)$ in the cut
$g$ plane: Bender-Wu singularities are confined in the angular region,
$\pi<{\rm Arg}\,g<2\pi$; $|g|<g_0$ (and a similar region with $g\to-g$),
but no further details on their location have been used. A tighter
bound (and a wider convergence region in $g$) can be obtained by
a further deformation of the contour near $\beta=1$, if $E(g)$ turns
out to be actually analytic in a wider region. Indeed, if $E(g)$ would
be analytic in the region, $|{\rm Arg}\,g|<\pi+\Delta$,
$0<\Delta<\pi/2$, then $1/(2g_0^{2/3})$ in Eqs.~(\ref{newboundB}),
(\ref{complexg}), (\ref{complexg'}) would be replaced by
$\cos[(\pi-2\Delta)/3]/g_0^{2/3}$. The estimate of $g_0$ from
our numerical study of delta expansion for the double well,
would accordingly be changed to $0.10\le g_0\le0.42$, which seems to be
consistent with the known strong coupling expansion coefficients
\cite{wjanke,bonnier}.
\item{(v)}
$E(e^{i\pi}g)$ with real positive $g$ corresponds to the energy
in an unstable potential well. According to our result, the complex
energy ``eigenvalues'' of such a system can be correctly reconstructed
by delta expansion/order dependent mapping, as long as $g>g_0$
(sliding regime, rather than tunnelling). This nicely explains the
empirical success of a variational-perturbative approach discussed
in Ref.~\cite{sliding}.

\ede

\subsection{General sufficient conditions for convergence}
\label{sec:proof3}
In this subsection the (sufficient) conditions for the convergence
of DE-ODM are summarized in the form of three theorems.

\medskip
{\bf Theorem 1:}
{\it Let a function $S(g)$ be given such that:
\bede
\item{1)}
$S(g)$ is analytic in the complex $g$ plane cut along negative axis,
continuous on the two edges of the cut and bounded at $g=0$ (on this sheet);
\item{2)}
$S(g)\sim g^p$ with $p>0$ for $g\to\infty$ on the cut plane;
\item{3)}
The discontinuity of $S(g)$ along the cut behaves as
\beq
   {\rm Disc}\,S(g)\sim C_S(-g)^{-c}\exp\left[-{a\over(-g)^{1/b}}\right]
\label{newdisc}
\eeq
for $g\to0^-$ with $a>0$, $1\le b<2$, $c$ real,
\ede
then the sequence $\{S_N(g)\}$ of approximants for $S(g)$, constructed
with conformal transformation
\beq
   g=\rho F(\beta)=\rho{\beta\over(1-\beta)^\alpha};
   \quad(\rho{\rm\ real\ positive})
\label{confor1}
\eeq
with
\beq
   1<\alpha\leq2,
\label{newalpha}
\eeq
followed by a Taylor expansion in $\beta$ up to order $N$ of
\beq
\label{generalpsi}
   \Psi(\beta,\rho)\equiv(1-\beta)^\sigma S(g(\beta))
\label{genpsi}
\eeq
with
\beq
   \sigma\ge\alpha p,
\label{newsigma}
\eeq
converges to $S(g)$ as $N\to\infty$ uniformly in every compact subset of
the region
\beq
   {\rm Re}\,{1\over g^{1/\alpha}}>0,
\label{gco1}
\eeq
if the real positive parameter $\rho$ is chosen to scale with $N$ as
\beq
  \rho={C'\over N^{\gamma'}};\quad{\rm with}\quad b<\gamma'<\alpha.
\label{mucondition}
\eeq
($C'$ independent of $g$.)
}

\medskip
{\bf Theorem 2:}
{\it Let $S(g)$ be given which satisfies 1)--3) of Theorem 1 and moreover:
\bede
\item{2')} its large $g$ expansion
\beq
   S(g)=g^p\sum_{n=0}^\infty d_n(g^{-q})^n
\label{scoupling}
\eeq
converges uniformly for $|g|\ge g_0$ and there exists a strictly
positive integer $k$ such that
\beq
   q<k<2q.
\label{kcondition}
\eeq
\ede
Then the sequence $\{S_N(g)/g^p\}$ with $S_N(g)$ constructed as in
Theorem~1 with the conformal transformation Eq.~(\ref{confor1})
with $\alpha=k/q$ (thus lying in the range $1<\alpha<2$), by use of the
function $\Psi(\beta,\rho)$ with $\sigma=\alpha p$, converges to
$S(g)/g^p$ as $N\to\infty$, uniformly in each closed subset of the
domain (including $g=\infty$)
\beq
   {\rm Re}\,{1\over g^{1/\alpha}}+
   {1\over g_0^{1/\alpha}}\cos\left[\pi(1-1/\alpha)\right]>0
\label{hearthburn}
\eeq
for any choice of scaling Eq.~(\ref{mucondition}) of the positive parameter
$\rho=C'/N^{\gamma'}$. ($C'$ independent of $g$.)
}

\medskip
{\bf Proof of Theorem 1:}
To prove Theorem 1 one proceeds as in
Sec.~\ref{sec:proof}--\ref{sec:proof1}: using {\it 1)\/} and {\it 2)\/}
one can write a dispersion relation as in Eq.~(\ref{cauchy}) for
$\Psi$ from which the expression Eq.~(\ref{remainder}) for the
remainder follows. In particular from {\it 1)\/} and {\it 2)\/}
one derives $|S(g)|\le A+B|g|^p$ in the cut plane including the edges of
the cut along the negative real axis, hence (with the choice
Eq.~(\ref{newsigma})) $\Psi$ is bounded by a constant on
$\Gamma\cup{\cal D}$ ($\Gamma$ and $\cal D$ are defined in
Sec.~\ref{sec:proof}). The integration contour $\Gamma$ is then split
in $\Gamma_A$ and $\Gamma_B$ as in Sec.~\ref{sec:proof1}.

As for $R_N^{(A)}$, one expresses the discontinuity of $\Psi$ along
$\beta'\in[\beta_c,0]$ by that of $S(g(\beta'))$, obtaining
\beq
   R_N^{(A)}=
   {\beta^{N+1}\over(1-\beta)^\sigma}
   \int_{\beta_c}^0{d\beta'\over2\pi i}
   {(1-\beta')^{\sigma}{\rm Disc}\,S(g(\beta'))\over
    (\beta')^{N+1}(\beta'-\beta)}.
\label{RA}
\eeq
Then by using Eq.~(\ref{newdisc}) for ${\rm Disc}\,S(g)$
(remember that $g(\beta')\to0^-$ as $N\to\infty$ for $\beta'\in[\beta_c,0]$),
the inequality
\beq
   |\beta-\beta'|\ge{\rm const.}\,|\beta|,
\label{strictg'}
\eeq
(which is valid for $g$ in (\ref{gco1})) and the fact that the inequalities
\bea
   &&1\le|1-\beta'|\le{\rm const.},
\non\\
   &&{\rm const.}\,\rho|\beta'|\le|g(\beta')|\le\rho|\beta'|
\label{gbound'}
\eea
hold (uniformly) on $ \Gamma_A$, one finds an upper bound as was done in
Eq.~(\ref{integral}). Extending the integration range up to $\beta'=\infty$
one gets (on the chosen compact):
\beq
   |R_N^{(A)}|\le
   {\rm const.}\,{|\beta|^N\over|1-\beta|^\sigma}\rho^{-c}
   \left({\rho\over a^b}\right)^{N+c}\Gamma(b(N+c))
   \sim|g|^{\sigma/\alpha}(\rho N^b)^N.
\label{newnewboundA}
\eeq

As for $\Gamma_B$ contribution we observe that the choice of $\alpha$
Eq.~(\ref{newalpha}) is essential to derive the inequality
(see \ref{sec:properties}),
\beq
   |\beta'|\ge1.
\eeq
Also, for each compact subset of the domain Eq.~(\ref{gco1})
the inequality
\beq
   |\beta-\beta'|\ge{\rm const.}\,|1-\beta|
\eeq
holds (uniformly in $\beta$, $\beta'$). One immediately gets
(exploiting the boundedness of $\Psi$)
\beq
   |R_N^{(B)}|\le{\rm const.}\,{|\beta|^{(N+1)}\over|1-\beta|^{1+\sigma}}
   \sim\exp\left[-N{\rm Re}\,\left({\rho\over g}\right)^{1/\alpha}\right].
\label{newboundb}
\eeq

Theorem 1 easily follows from Eqs.~(\ref{newnewboundA}), (\ref{newboundb})
and (\ref{mucondition}) (note that the upper limit on $\gamma'$ comes
{}from $R_N^{(B)}$ while the lower limit from $R_N^{(A)}$).
{\it Q.E.D.}

\medskip
{\bf Proof of Theorem 2}
The new hypothesis {\it 2')\/} on the existence of a large $g$ expansion
of $S(g)$, together with the choices $\alpha q=k$ and $\sigma=\alpha p$,
ensures (see footnote \ref{weierstrass}) the analyticity of $\Psi$ in the
extended region ${\cal D}\cup{\cal C}$ enclosed by $\Gamma'$.

As in Sec.~\ref{sec:proof2}, the analyticity on the finite disk
$\cal C$ around $\beta=1$ together with the uniqueness of analytic
continuation implies that the function $\Psi(\beta,\rho)$ takes
the same values on all the ``mirror'' Riemann sheets, i.e., those
connected by going around the branch point at $\beta=1$. As a result,
on applying Cauchy's theorem one can work on a single sheet
corresponding to the given value of $g$; loops on all other
Riemann sheets give vanishing contribution, enclosing no poles.

We proceed as in Sec.~\ref{sec:proof2}, by splitting $\Gamma'$ in
$\Gamma_A$ and $\Gamma'_B$. The treatment of $\Gamma_A$ contribution is the
same as in Theorem 1 and one finds Eq.~(\ref{newnewboundA}).

As for the contribution of $\Gamma'_B$, we observe that the condition
Eq.~(\ref{kcondition}) (enforcing $1<\alpha<2$) and the chosen range
of $g$, Eq.~({\ref{hearthburn}), lead to the inequalities
\bea
   &&|\beta'-\beta|\ge\epsilon r_0;
\non\\
   &&|\beta'|\ge1+r_0\cos(\phi_\alpha)+O(r_0^2)
\non
\eea
for $\beta'\in\Gamma'_B$. ($\epsilon$ is a positive $\beta$, $\beta'$
independent fixed constant.) Together with the boundedness of $\Psi$
(which follows from {\it 2')}), these inequalities give, as in
Sec.~\ref{sec:proof2},
\bea
   |R_N^{(B)}|&\le&
   {\rm const.}\,{|\beta|^{N+1}\over|1-\beta|^\sigma r_0}
   \left({1\over1+r_0\cos\phi_\alpha}\right)^{N+1}
\non\\
   &\sim&|g|^p\exp\left[
    -N\rho^{1/\alpha}
    \left({\rm Re}\,{1\over g^{1/\alpha}}
             +{1\over g_0^{1/\alpha}}\cos\phi_\alpha\right)\right].
\label{newnewboundB}
\eea
The theorem follows from Eq.~(\ref{newnewboundA})
and Eq.~(\ref{newnewboundB}).
{\it Q.E.D.}

\medskip
The convergence proof expressed in the two precedent theorems actually
holds if the trial parameter $\rho$ (or $x$) is scaled with the order
$N$ precisely as $\rho\sim N^{-1}$ (or $x\sim N^{1/(2\alpha)}$), under
certain conditions on the proportionality coefficient. Since this case
was discussed earlier by several authors in connection with the optimized
DE-ODM for the anharmonic oscillator, we state the result for this
extreme case as a separate theorem.\footnote{
Note however that in general such a particular scaling of $\rho$
or $x$ does not follow from the optimization condition alone
(either principle of minimum sensitivity or fastest apparent convergence)
used frequently in the literature. It is only after imposing a further
condition, e.g., of choosing the root of smallest modulus that
optimization equation fixes the particular scaling of $x$.
See the related discussion in Sec.~\ref{sec:opti}.}

\medskip
{\bf Theorem 3:}
{\it Theorem 1 and Theorem 2 hold if (maintaining the other respective
hypotheses) we scale the parameter $\rho$, instead of
Eq.~(\ref{mucondition}) as
\beq
   \rho={C'\over N^b}
\label{extremalscaling}
\eeq
and
\beq
   0<C'<\left({a\over u_*}\right)^b
\eeq
where $u_*$ is a function of $\alpha$, $b$ defined in Eq.~(\ref{phieq2})
below (see also Eq.~(\ref{roughu}) for less stricter but $\alpha$,
$b$ independent range of convergence) and $a$, $b$ are those of
Eq.~(\ref{newdisc}).}

\medskip
{\bf Proof:}
We proceed exactly as in Theorem 1 and Theorem 2, splitting the
remainder in two pieces $R_N^{(A)}$ and $R_N^{(B)}$. With the scaling
Eq.~(\ref{extremalscaling}) the bounds on $R_N^{(B)}$
Eqs.~(\ref{newboundb}) and (\ref{newnewboundB}) still hold unmodified
and enforce convergence of that contribution in the corresponding range
of $g$. It will be shown that with the assumed scaling the contribution
$R_N^{(A)}$ can be bounded by an exponentially vanishing term
that is independent on $g$; all the restrictions on the range of $g$
come from $R_N^{(B)}$, accordingly they are the same as in the
previous theorems.

It is clear that with scaling Eq.~(\ref{extremalscaling}) the previous
bound Eq.~(\ref{newnewboundA}) is too loose to prove the convergence:
we thus step back to Eq.~(\ref{RA}) that is common to both theorems and
try to get a better bound. By use of asymptotic behavior
Eq.~(\ref{newdisc}) of ${\rm Disc}\,S(g)$, of bounds
Eq.~(\ref{strictg'}), Eq.~(\ref{gbound'}), extending the integral
range to $\beta'\in[-\infty,0]$ and changing the variable as
$\beta'\to-\beta'$, we find:
\begin{equation}
   |R_N^{(A)}|\le{\rm const.}\,{|\beta|^{N+1}\over|1-\beta|^\sigma}
   \rho^{-c}\int_0^\infty d\beta'
   \exp
   \left[-(N+1+c)\Phi\left(\beta',{a\over(N+1+c)\rho^{1/b}}\right)\right],
\label{extremalbound}
\end{equation}
where
\begin{equation}
   \Phi(\beta',u)\equiv\log\beta'
   +u{(1+\beta')^{\alpha/b}\over{\beta'}^{1/b}}.
\label{action}
\end{equation}
Eq.~(\ref{extremalbound}) can be estimated by the saddle point method.
For fixed $u>0$, $\Phi$ behaves as $u/\beta'^{1/b}$ near $\beta'=0$ and as
$u\beta'^{(\alpha-1)/b}$ for $\beta'\to\infty$ thus at least one saddle point
(depending on $u$ and others parameters $\alpha$, $b$), $\beta'_*(u)$
exists, defined by
\beq
   \de_{\beta'}\Phi(\beta'_*,u)=0.
\label{phieq1}
\eeq
If (depending on $\alpha$, $b$) there are more than one such point, the
one corresponding to smallest $\Phi(\beta'_*,u)$ must be chosen. The
bound for $R_N^{(A)}$ is then
\bea
   |R_N^{(A)}|&\le&
   {\rm const.}\,{|\beta|^{N+1}\over|1-\beta|^\sigma}\rho^{-c}
   {1\over\sqrt{\de^2_{\beta'}\Phi(\beta'_*,u)}}
   e^{-{(N+1+c)\Phi(\beta'_*,u)}}
\non\\
\label{extremeboundb}
   &\sim&e^{-{N\Phi(\beta'_*,u)}}
\eea
(in the last line powers of $N$ have been neglected). The function
$\Phi_*(u)\equiv\Phi(\beta'_*(u),u)$ is an increasing function of $u$,
as can be easily seen from (due to Eq.~(\ref{phieq1}))
\beq
   {d\Phi_*\over du}=\de_u\Phi(\beta'_*, u)+
   \de_{\beta'}\Phi(\beta'_*,u){\de \beta'_*\over\de u}=
   \de_u\Phi(\beta'_*,u)>0
\eeq
(last inequality comes from a direct inspection of Eq.~(\ref{action})).
Thus if we restrict ourselves to $u>u_*$, with $u_*$ defined by
\beq
   \Phi_*(u_*)=\Phi(\beta'_*(u_*),u_*)=0,
\label{phieq2}
\eeq
the bound on $R_N^{(A)}$ vanishes exponentially.
{\it Q.E.D.}

\medskip
The value of $u_*$ in Theorem 3 should be obtained (for fixed $\alpha$, $b$)
by numerical solution of coupled equations (\ref{phieq1}) and (\ref{phieq2}).
Nevertheless it is easy to get a (more conservative) $\alpha$, $b$
independent range of convergence in terms of $u$. Indeed, if we bound
$\Phi$ in Eq.~(\ref{action}) with
\beq
   \Phi(\beta',u)\ge\widetilde\Phi(\beta',u)\equiv\log\beta'
   +{u\over{\beta'}^{1/b}}
\eeq
and proceed as in Theorem 3, the equations analogous to Eq.~(\ref{phieq1})
and Eq.~(\ref{phieq2}) can be solved exactly, obtaining the
exponential convergence ($\widetilde\Phi(\beta'_*,u)>0$) for
\beq
   u>be^{-1}\ge e^{-1}.
\label{roughu}
\eeq

\section{Convergence of the Optimized (Variational) Expansion}
\label{sec:opti}
Delta expansion - order dependent mappings {\it diverges\/} if
the parameters $\rho$ or $x$ depends on $N$ improperly, i.e.,
if they lie outside the convergence range, Eqs.~(\ref{gamma}),
(\ref{gamma'}), (\ref{extremex}), (\ref{extreme}). In the case where $x$
grows with $N$ faster than $N^{1/2}$, the behavior of the remainder
$R_N$ can be estimated by observing that its main contribution
comes from the integration region near $\beta'=1$ on $\Gamma'_B$. One finds
\beq
   S_N\sim \left({x^2\over N}\right)^\sigma,
\eeq
(note that $x/N^{1/2}\to\infty$; $\beta^N\to1$). For the particular
case $\sigma=1/2$ it reduces to the result found earlier \cite{GKS}.

On the other hand, if $x$ grows too slowly ($x/N^{1/2\alpha}\to0$),
the main contribution to $R_N$ comes from the region $\beta'\sim0^-$.
This contribution can be easily estimated by the saddle point
approximation and gives:
\beq
   S_N\sim\left(-{N\over x^{2\alpha}}\right)^N
\eeq
(this behavior is an offspring of the standard divergence of
perturbative coefficients).

It is clear that there are no solutions of optimization condition
(e.g., $\partial S_N/\partial\Omega=0$, or $S_{N+1}-S_N=0$) in these ranges
of $x\equiv\om/\omega$. Optimization forces the parameter $x$ (or $\rho$)
to lie within the convergence range. A study of the roots of
optimization equation (see Appendix C) with regard to the $g$ dependence
of $\rho$ or $x$, gives a further consistency check of this point.

In other words, we have proven that {\it optimized\/} DE-ODM for
anharmonic oscillator/double well system converges as long as the
coupling constant lies in the appropriate domain. This is important
since in many (especially in higher dimensional) systems it is a hard
task to do higher order calculations. Optimization is crucial to get
a good approximation from lowest order calculations.

Of course, optimization condition alone does not usually yield a unique
solution for $x$ (or $\rho$), or a given optimization criterion may not
have any real solution at all at a given order $N$. Furthermore, as the
number of the roots of optimization equation in general grows with $N$,
it is not very practical to focus our attention to a particular root and
to try to follow it: it may disappear as $N$ is varied and a new pair of
roots may appear, etc.\ \cite{DJ}.

In the case of the energy eigenvalues of anharmonic oscillator
or double well, a general criterion that follows from our analysis
is that the convergence rate is fastest for the smallest possible
$\om$, in accordance with the earlier suggestions \cite{SZ}.

In any event, the importance of our results in the precedent
sections lies in the fact that they do prove the convergence of
optimized or variational expansion, without however letting
ourselves involved into the analysis of complicated behavior of
the solutions of a particular optimization scheme. We find it remarkable
that a variational method is proven to converge, even if no variational
principle is involved here.

\section{$q^{2M}$ oscillators with $M\ge 3$}
\label{sec:q2M}
The convergence proofs presented in Sec.~\ref{sec:proof} do not apply
directly to the case of higher anharmonic oscillators ($M\ge3$ in
Eq.~(\ref{aho})). For simplicity consider the case of real $g$.
All the pre-requisites of Theorem 1 are satisfied:
$E(g)$ is known to be analytic in the cut $g$ plane \cite{LM,Simon},
and the conditions {\it 2)\/} and {\it 3)\/} are satisfied with
$p=1/(M+1)$ and $b=M-1$. The convergence cannot however be proven
in these cases because there are no values of the scaling index $\gamma'$
satisfying $b<\gamma'<\alpha$, since $b=M-1\ge2$ on the one hand, and
$1<\alpha\le2$ on the other.

This does not yet mean that DE-ODM diverges in these cases, since
Theorem 1 gives only sufficient conditions for convergence. Let us
re-examine the last condition ($\alpha\le2$) imposed on the conformal
transformation. It is needed so that the complex part ($\Gamma_B$) of the
image of the negative real $g$ axis lies outside of the unit circle, i.e.,
$|\beta'|>1$, $\forall\beta'\in\Gamma_B$, $\beta'\ne1$.

Now one knows that $E(g)$ is analytic on the cut $g$ plane, hence
$E(g(\beta))$ analytic within $\cal D$ surrounded by $\Gamma_A$, $\Gamma_B$.
If the function $\Psi(\beta,\rho)=(1-\beta)^\sigma E(g)$ would be
analytic in a slightly (but sufficiently) larger region of $\beta$,
one could consider a conformal transformation with $\alpha>2$
(actually one needs $\alpha>M-1$), deform (enlarge) the integration
contour $\Gamma_A$, $\Gamma_B$ to $\Gamma'_A$, $\Gamma'_B$ until
$|\beta'|>1$, $\forall\beta'\in\Gamma'_B$. All the steps of the
convergence proof would then be valid. Note however that such a
deformation in the $\beta$ plane involves going onto the higher Riemann
sheets in the original $g$ plane. It is here that Bender-Wu type
singularities potentially make obstructions.
Unfortunately, having no detailed knowledge on the positions of
Bender-Wu type singularities in the general $q^{2M}$ oscillators,
we are at present unable to answer whether or not DE-ODM for $q^{2M}$
($M\ge3$) oscillators converges with an appropriate scaling of $\rho$
(or by optimization).

It is interesting that Pad\'e's method does not converge \cite{Grecchi}
for $q^{2M}$ oscillators, for $M>3$, either.

\section{Green's functions in the one dimensional $\phi^4$ theory}
\label{sec:green}
\def\arguments{\{{\omega}{\tau_i}\}}
As an application of Theorem 1 of Sec.~\ref{sec:proof3}, consider the
{\it Euclidean\/} Green's functions of the anharmonic oscillator system:
\bea
   &&\left\langle q(\tau_1)q(\tau_2)\cdots q(\tau_n)\right\rangle
\non\\
   &&={1\over Z}\int Dq\,q(\tau_1)q(\tau_2)\cdots q(\tau_n)
   \exp\left[-\int_{-\infty}^\infty d\tau
   \left({\dot q^2\over2}+{\omega^2\over2}q^2+{g\over4}q^4\right)\right]
\non\\
   &&\equiv{1\over\omega^{n/2}}
   G_n\left({g\over\omega^3};
            \omega\tau_1,\omega\tau_2,\cdots\omega\tau_n\right).
\label{green}
\eea
A simple-minded application of the delta-expansion based on the shift
of the quadratic term of the Hamiltonian Eq.~(\ref{sub}), seems to be
problematic, due to the complicated dependence on $\omega\tau_i$.
One may, however, well apply Theorem 1 directly to
$G_n(\widetilde g,\arguments)|_{\tilde g=g/\omega^3}$ treating
$\arguments$ as fixed parameters.

We first assume the analyticity of $G_n(g,\arguments)$ in the whole cut
plane of $g$. This is quite reasonable, since any given Green function
can be expressed as an (infinite) sum of terms analytic in the cut
plane, by expanding in the energy eigenstate basis. Such an assumption
was also made by Strharsky \cite{strh} to compute successfully the large
order behavior of the Euclidean Green's functions by use of the
dispersion relation. The other conditions {\it 2)\/} and {\it 3)\/}
of Theorem 1 are satisfied, with $p=-n/4$; $a=4/3$; $b=1$; $c=n/2$.

DE-ODM for $G_n$ constructed from its standard perturbation series
by the conformal transformation Eq.~(\ref{confor1}), followed by the
Taylor expansion in $\beta$, then converges to the correct answer, if
$\rho$ is scaled with the order $N$ as in Eqs.~(\ref{gamma}), (\ref{mu})
or Eqs.~(\ref{extremex}), (\ref{extremeC}).

The convergence proof above holds {\it pointwise\/} for each fixed value
of $\arguments$. In fact a {\it uniform\/} convergence over arbitrary
value of $\arguments$ is desired, because the functional form of $G_n$,
$\arguments$ as the argument, is what one wishes to reproduce.
The following argument can be given on this point using the result of
\cite{GKS}.

If we apply the same procedure as in \cite{GKS} also for $G_n$,
we have an expression for the remainder similar to Eq.~(2.53) of \cite{GKS}.
A part (corresponding to $R_N^{(B)}$ above) of the remainder is bounded
(essentially) by $c_1\beta^N$, where $c_1$ is the first order
perturbative coefficients for $G_n$. Another part (corresponding to
$R_N^{(A)}$ above) is estimated by the Euclidean bounce solution for
$g<0$. Now $c_1$ is computed by using the free propagator,
$\langle q(\tau)q(\tau')\rangle_0=e^{-\omega|\tau-\tau'|}/(2\omega)$,
thus obviously is bounded for the whole range of $\arguments$. On
the other hand, $\arguments$ dependence of $R_N^{(A)}$ is
proportional to (for $N$ large)
$\int_{-\infty}^\infty d\tau_0\prod_i\sqrt{2}
{\rm sech}\,\omega(\tau_i-\tau_0)$,
where $\tau_0$ is the position of the bounce center, and this is also
bounded for arbitrary value of $\arguments$. This argues strongly
for the uniform convergence with respect to $\arguments$.

\section{Summary and discussion}
\label{sec:summary}
Delta expansion - order dependent mappings, is a powerful resummation
technique which converts a given divergent series, under certain
conditions, into a convergent sequence of approximants. For its
generality, the extreme simplicity of use (in particular,
combined with an optimization procedure), and for its impressive success
in simple quantum mechanical systems, it seems to be worthwhile to
study carefully the mechanism underlying such a method.

In this paper, the convergence proof for the DE-ODM given previously
by the present authors for the anharmonic oscillator has been refined
and its domain of applicability considerably extended. The key step
for the improvement was the use of Cauchy's theorem, written directly
in the complex plane of the conformally transformed variable, instead
of making a conformal transformation in the standard dispersion
relation. Use of the knowledge on the analyticity property of $E(g)$
on the full Riemann surface was also crucial.

In particular, this allowed us to show the uniformity of convergence
in a wide region of the Riemann surface (Fig.~3), including
the strong coupling limit of anharmonic oscillator ($g=\infty$),
and the large $g$ regime of the quantum mechanical double well.

It is interesting to note that in this last case the standard
perturbation series is non Borel summable. Together with the example of
the zero dimensional model discussed in \ref{sec:zero}, our results
convincingly demonstrate the fact that the non Borel summability of the
standard perturbation series is no obstruction to the existence of
a simple resummation procedure which converges to the exact answer.
The possibility of reconstruction in non Borel summable cases here
hinges upon the known existence of a wide region of analyticity of
$E(g)$, which is larger than normally required for the Borel summability
\cite{sokal}.

As for the variational aspect of the delta expansion - order dependent
mappings, it is of considerable interest that a variational method
can be proven to converge, without use of the variational principle.

A lesson to be learned from the results found here is the fact that,
if only convergence for real finite $g$ is required, quite a wide
arbitrariness exits in the choice of the conformal transformation,
to be used in the construction of DE-ODM. Our convergence proof can
probably be generalized to conformal transformations of more general forms.

At the same time, we have clearly marked the limit of validity of the
DE-ODM based on a simple conformal transformation of $g$. In particular,
the perturbation series of the quantum mechanical double well in the
regime of weak coupling (where tunnelling effects---or instantons---are
clearly essential), cannot be resummed by the present method.

Also, in four dimensional field theory models such as Quantum Chromodynamics,
where the ana\-ly\-tic struc\-ture in the re\-norma\-lized
coupl\-ing cons\-tant is far more complicated~\cite{thooft},
the validity of an analogous approach is not obvious.

Nevertheless, we believe that our general theorems apply to many
quantum mechanical and field theory models in dimensions less than four.

\noindent
{\bf Note Added}

When this paper was being typed the authors were informed that an
analogous technique was used by H. Kleinert \cite{PC} to show
the convergence of DE-ODM in the strong coupling limit.

\noindent
{\bf Acknowledgments}
\nopagebreak

One of the authors (K.K.) thanks the members of the Laboratoire de
Physique Th\'eorique et Hautes Energies, Centre d'Orsay, Universit\'e
de Paris Sud, for a warm hospitality extended to him. We thank
J. Zinn-Justin, J. Maharana and S. Rudaz for discussions, and
Andr\'e  Martin, T. T. Wu and H. Kleinert for useful information.
The work of R.G. is partially supported by Italian Ministry of University
and of Scientific and Technological Research (MURST).
The work of H.S. is supported in part by Monbusho Grant-in-Aid
Scientific Research No.~07740199 and No.~07304029.

\appendix
\section{Some properties of the conformal transformation}
\label{sec:properties}
We study here how the negative $g$ axis is mapped by the inverse of the
conformal transformation
\beq
   g=\rho{\beta\over(1-\beta)^\alpha}
\label{NNN}
\eeq
for different values of $\alpha>1$. The problem is equivalent to the study
of solutions of the equation
\beq
   H(\beta,s)\equiv(1-\beta)^\alpha+s\beta=0
\label{equation}
\eeq
with $s=-\rho/g >0$.

For $s>s_c\equiv\alpha^\alpha/(\alpha-1)^{\alpha-1}$
Eq.~(\ref{equation}) has two real solutions that coalesce
(one coming from $\beta=-\infty$, the other from $\beta=0^-$) for
$s=s_c$ at $\beta=\beta_c\equiv-1/(\alpha-1)$. These values are
found by imposing $H(\beta_c,s_c)=\de_\beta H(\beta_c,s_c)=0$. For
$0<s<s_c$ the equation (\ref{equation}) has two complex conjugate
solutions that start from $\beta=\beta_c$ and join at $\beta=1$.

Other complex conjugate solutions converging to $\beta=1$ at $s=0$
are in general also present,\footnote{
It can be shown that there exist $2(k+1)$ roots on the first Riemann
sheet of the mapping Eq.~(\ref{NNN}), $-\pi<{\rm Arg}\,(1-\beta)<\pi$,
in which $k$ is a nonnegative integer such that $2k+1\leq\alpha<2k+3$.
For the case of our main interest, $1<\alpha\leq2$, therefore only
two roots exist on this sheet.}
but the complex contour that wraps around the cut $g$ plane is mapped
to the contour $\Gamma$ formed by the two above mentioned complex
conjugate paths (called $\Gamma_B$ in Sec.~\ref{sec:proof}) plus the
path $[\beta_c,0]$ ($\Gamma_A$) of the smaller negative root.

First of all, solving Eq.~(\ref{equation}) in the limit $s\to0$
(i.e., $\beta\sim1$) one easily finds that $\Gamma_B$ leaves the point
$\beta=1$ with angle $\phi_\alpha =\pm(1-1/\alpha)\pi$.

To study better the shape of $\Gamma_B$, set $\beta=|\beta|e^{i\delta}$
and consider $|\beta|$ at a fixed phase~$\delta$, $0<\delta<\pi$,
as a function of $\alpha$. One finds from Eq.~(\ref{equation}),
after eliminating $s$,
\beq
   |\beta|=\sin{\pi-\delta\over\alpha}\bigg/
           \sin{(\alpha-1)(\pi-\delta)\over\alpha}.
\eeq
It is then easy to see that
\bea
   &&|\beta|>1,\quad{\rm if}\quad 1<\alpha<2;
\non\\
   &&|\beta|<1,\quad{\rm if}\quad\alpha>2,
\eea
for $\forall\delta$, $0<\delta<\pi$.

\section{Zero dimensional model}
\label{sec:zero}
In this appendix, we consider the delta expansion or, equivalently,
the order dependent mapping with conformal transformation,
\beq
   g=\rho\beta/(1-\beta)^2,
\label{zero}
\eeq
of a simple integral,
\begin{equation}
   Z(g)={1\over\sqrt{\pi}}\int_{-\infty}^\infty dq\,e^{-q^2-gq^4}
\label{ichi}
\end{equation}
(the partition function of $\phi_0^4$ theory).
Due to the absence of Bender-Wu type singularities, DE-ODM in this
case converges for each $g$ in the {\it whole Riemann surface},
in the limit, $N\to\infty$, $\rho=C'/N^{\gamma'}\to0$ with
\bea
   &&1<{\gamma'}<2,\quad{\rm or}
\non\\
   &&{\gamma'}=1;\quad C'<C'_{\rm crit.}=
   1.116711531873221\cdots.
\label{jyuu}
\eea
(In terms of $x$, $x\sim(g/\rho)^{1/4}\to\infty$.)\footnote{
Previously convergence was proved only pointwise for real $g$
\cite{SZ,buck,GKS} and for a particular phase of $g$, ${\rm Arg}\,g=2\pi$
(corresponding to the double well) \cite{DJ}.}

The proof for $g$ lying on higher Riemann sheets in this case requires
a non-trivial extension of the treatment given in Sec.~\ref{sec:proof}.

We start with the observation that the analytic continuation of $Z(g)$
is given in terms of the parabolic cylinder function by
\beq
   Z(g)={1\over(2g)^{1/4}}e^{1/(8g)}D_{-1/2}(1/\sqrt{2g})
\label{ni}
\eeq
with the strong coupling expansion
\beq
   Z(g)={1\over2\sqrt{\pi}}g^{-1/4}\sum_{n=0}^\infty
   (-1)^n{\Gamma(n/2+1/4)\over n!}(g^{-1/2})^n
\label{san}
\eeq
(the latter can easily be derived directly from Eq.~(\ref{ichi})).

Following the general procedure described in Sec.~\ref{sec:proof2} we
introduce the function $\Psi(\beta,\rho)$ by
\beq
   Z(g)\equiv(1-\beta)^{1/2}\Psi(\beta,\rho),
\eeq
where the $N$th order approximant for $Z(g)$ is to be constructed as
$Z_N(g)\equiv(1-\beta)^{1/2}\cdot\left\{\Psi(\beta,\rho)\right\}_N$;
$\{f(\beta)\}_k$ is the $k$th order truncation of the Taylor expansion of
$f(\beta)$ in $\beta$.

In order to write Cauchy's theorem in the $\beta$ variable, the analytic
property of $\Psi$ must be studied first. A good knowledge of it can be
deduced from the strong coupling expansion Eq.~(\ref{san}). Indeed,
it is easily shown that this expansion converges for any $g\ne0$,
implying that the function $\Psi$ defined above is analytic everywhere,
except at the quartic branch points at $\beta=0$ and at $\beta=\infty$.
($\beta=0$ is also an essential singularity, see Eq.~(\ref{nijyuushi})
below.)

The image of the cut $g$ plane is the interior of a unit circle (minus
the negative real axis) in the first $\beta$ Riemann sheet and
it is tempting to take its boundary as the integration contour
$\Gamma=\Gamma_A+\Gamma_B$ and make a ``small'' enlargement of
the contour around $\beta=1$, keeping the rest of the contour
unchanged, as was done in Sec.~\ref{sec:proof2}. It turns out that
this procedure does not work here. (The convergence factor
$1/(1+r_0\cos\phi_\alpha+O(r_0^2))^N$ in Eq.~(\ref{newnewboundB})
is lost in the case $\alpha=2$.) On the other hand, the analyticity
domain of $\Psi$ is much wider here, so that the contour can be
enlarged more substantially. Such enlargement is indeed possible but
involves a delicate issue; it can be dealt with as follows.

We write the remainder $R_N(\beta)\equiv Z(g)-Z_N(g)$ as:
\begin{equation}
   R_N(\beta)=(1-\beta)^{1/2}
   \beta^{N+1}\oint_{\Gamma'_A+\Gamma'_B}{d\beta'\over2\pi i}
   {\Psi(\beta')\over(\beta')^{N+1}(\beta'-\beta)},
\label{jyuuku}
\end{equation}
where $\Gamma'_B$ will be taken as a circle with the radius $R>1$
and $\Gamma'_A$ as a contour which wraps around a part of the negative
real axis $-R\le\beta'\le0$ (see Fig.~4).

The radius $R$ will be taken dependent on the order $N$ as
\begin{equation}
   R=1+dN^{-\kappa},\quad(d>0;\quad{\gamma'}/2>\kappa>0).
\label{jyuuroku}
\end{equation}
In deriving Eq.~(\ref{jyuuku}), it was implicitly assumed that
$|\beta|<R$. This indeed holds for large $N$ with this choice of $R$.
($\beta\sim1-(\rho/g)^{1/2}$ as $\rho\to0$.)
(For the subtleties involved in writing Cauchy's theorem on $\beta$
Riemann surface and for its elucidation, see the comment before
Eq.~(\ref{qui}) in the main text.)

\smallskip
On $\Gamma'_B$, $|\beta'|=R$ and $|\beta'-\beta|\geq
R-|\beta|$, hence the contribution to $R_N(\beta)$ is bounded by
\begin{equation}
   \left|R_N^{(B)}(\beta)\right|\leq
   |1-\beta|^{1/2}\left({|\beta|\over R}\right)^{N+1}
   {{\rm max}\,|\Psi(\beta')|\over1-|\beta|/R}.
\label{nijyuuichi}
\end{equation}
When $\beta'$ is on the contour $\Gamma'_B$, the corresponding
$g'=\rho\beta'/(1-\beta')^2$ circles around the origin $g'=0$
on the second Riemann sheet, with the radius,
\begin{equation}
   |g'|=\rho{R\over|1-\beta'|^2}\leq\rho{R\over(R-1)^2}.
\label{nijyuusan}
\end{equation}
With $\rho$ scaling as $\rho=C'/N^{\gamma'}$, and $R$ as Eq.~(\ref{jyuuroku}),
$|g'|$ is small for $N$ large. We can therefore use the asymptotic
expansion near the origin,\footnote{
The plus sign is for $-5\pi/2<{\rm Arg}\,g<-\pi/2$ and the minus sign is for
$\pi/2<{\rm Arg}\,g<5\pi/2$; the terms with the exponential factor is absent
in other regions of $g$ \cite[Eq.~(9.246)]{GR}.}
\begin{equation}
   Z(g)\sim{1\over\sqrt{\pi}}
   \sum_{n=0}^\infty{\Gamma(2n+1/2)\over n!}
   \left[(-1)^n\pm i\sqrt{2}e^{1/(4g)}\right]g^n,
\label{nijyuushi}
\end{equation}
to get the following bound on $\Psi(\beta')$:\footnote{
We have taken the dominant
term for ${\rm Re}\,g'>0$. When ${\rm Re}\,g'<0$, $|Z(g')|$ is bounded
by a power of $|g'|$ and the contribution is harmless for the
convergence. The existence of such a large factor near the origin
on the second Riemann sheet is understood as $g=e^{2\pi i}|g|$
corresponds to double well model.}
\begin{equation}
   |\Psi(\beta')|\leq
   {\sqrt{2}\over(R-1)^{1/2}}\exp\left[{(R-1)^2\over4\rho R}\right]
   (1+O(\rho R/(R-1)^2)).
\label{nijyuuroku}
\end{equation}
We get
\begin{eqnarray}
   \left|R_N^{(B)}(\beta)\right|
   &\leq&
   {\sqrt{2}|1-\beta|^{1/2}\over(1-|\beta|/R)(R-1)^{1/2}}
   \left({|\beta|\over R}\right)^{N+1}
   \exp\left[{(R-1)^2\over4\rho R}\right]
\nonumber\\
   &\sim&\exp\left(-dN^{1-\kappa}+{d^2\over4C'}N^{{\gamma'}-2\kappa}\right).
\label{nijyuushichi}
\end{eqnarray}
Note that for $0<\gamma'<2$, it is always possible to choose $\kappa$ in
such a way that
\begin{equation}
   0<\gamma'-2\kappa<1-\kappa.
\label{nijyuuhachi}
\end{equation}
With the first factor dominating, one gets $|R_N^{(B)}(\beta)|\to0$
as $N\to\infty$.

On $\Gamma'_A$, $-R\le\beta'\le0$. The value of $g'$ corresponding to it
moves along $-\rho/4\le g'\le0$. Since $\rho\ll1$ for $N$ large,
the asymptotic form
\begin{equation}
   |{\rm Im}\,\Psi(\beta')|=
   {|{\rm Im}\,Z(g')|\over|1-\beta'|^{1/2}}
   \sim{1\over\sqrt{2}}{e^{1/(4g')}\over|1-\beta'|^{1/2}}(1+O(g')),
\end{equation}
can be used in
\beq
   R_N^{(A)}(\beta)
   =(1-\beta)^{1/2}\beta^{N+1}\int_{-R}^0{d\beta'\over\pi}
   {{\rm Im}\,\Psi(\beta')\over(\beta')^{N+1}(\beta'-\beta)}.
\eeq
Therefore, using $|\beta'-\beta|\geq|\beta|$ and $|1-\beta'|\geq1$, we get
\begin{equation}
   \left|R_N^{(A)}(\beta)\right|
   \leq{1\over\sqrt{2}}|1-\beta|^{1/2}|\beta|^N\int_0^\infty
   {d\beta'\over\pi}e^{-(N+1)\Phi(\beta')}
   (1+O(1/N)),
\label{nijyuuku}
\end{equation}
where $\Phi(\beta')=\ln\beta'+u(1+\beta')^2/\beta'$, $u=1/(4(N+1)\rho)$
and in the last step we have changed the variable $\beta'\to-\beta'$
and extended the upper limit of integration to $\infty$.
The right hand side of Eq.~(\ref{nijyuuku}) can be evaluated by the saddle
point method for $N$ large, as was done in Sec.~\ref{sec:proof2},
Sec.~\ref{sec:proof3}. Combining the bound for $R_N^{(A)}$ thus obtained
with that for $R_N^{(B)}$, we get the announced result.

An interesting application of our result is the case of the
``double well'' model in zero dimension:
\begin{equation}
   \widetilde Z(g)={1\over \sqrt{\pi}}\int_{-\infty}^\infty dq\,e^{q^2-gq^4}
   ={1\over(2g)^{1/4}}e^{1/(8g)}D_{-1/2}(-1/\sqrt{2g}),
\end{equation}
with $g$ real positive. From the known properties of the parabolic
cylinder function one finds that
$\widetilde Z(g)=(e^{2\pi i})^{1/4}Z(e^{2\pi i}g)$
(an analogue of the Symanzik's scaling relation). Consequently,
the (scaled) delta expansion of the double well model
(whose standard perturbation series is non Borel summable) converges
to the exact answer for {\it any\/} real value of $g\ne0$, in accordance
with an earlier result \cite{buck,DJ}.

As pointed out in \cite{SZ}, the case ${\rm Arg}\,g=\pi$ can also be
related to $\widetilde Z(g)$. The crucial relation is
\begin{equation}
   \widetilde Z(g)=\sqrt{2}e^{1/(4g)}{\rm Re}\,Z(e^{\pi i}g).
\end{equation}
Again according to our result, the exact answer for $\widetilde Z(g)$
can be correctly reconstructed systematically by using
DE-ODM for $Z(e^{\pi i}g)$, whose standard perturbation series
$Z(e^{\pi i}g)\sim(1/\sqrt{\pi})\sum_{n=0}^\infty\Gamma(2n+1/2)g^n/n!$
is non Borel summable.

\section{Optimization condition on $\rho$}
\label{sec:optimization}
In this appendix, we show that either of optimization conditions,
the principle of minimal sensitivity (PMS) or the fastest apparent
convergence (FAC), leads a simple equation for $\rho$ which is
{\it independent\/} on $g$. This fact, for the particular case
with $\sigma=1/2$ and $\alpha=(M+1)/2$, has recently been noted by
\cite{Janke}.

By PMS, one imposes $dS_N^{\rm (ODM)}/dx=0$ (or equivalently
$dS_N^{\rm (ODM)}/d\rho=0$) order by order. From the definition of
DE-ODM, we know
\begin{eqnarray}
   S_N^{\rm(ODM)}&=&
   {1\over(1-\beta)^\sigma}\sum_{n=0}^Nc_n\rho^n
   \left\{\beta^n(1-\beta)^{-\alpha n+\sigma}\right\}_N
\nonumber\\
   &=&\sum_{n=0}^Nc_ng^n(1-\beta)^{\alpha n-\sigma}
   \left\{(1-\beta)^{-\alpha n+\sigma}\right\}_{N-n},
\end{eqnarray}
where $\{f(\beta)\}_k$ is a $k$th order truncation of the Taylor series
on $\beta$. With a use of a relation
\begin{equation}
   {d\over d\beta}
   \left[(1-\beta)^\ell\left\{(1-\beta)^{-\ell}\right\}_k\right]=
   -{\Gamma(\ell+k+1)\over k!\,\Gamma(\ell)}(1-\beta)^{\ell-1}\beta^k,
\end{equation}
we have (note the variation with respect to $x$ is taken while $g$
is kept fixed)
\begin{equation}
   {d\over dx}S_N^{\rm (ODM)}=
   -{d\beta\over dx}{\beta^N\over(1-\beta)^{\sigma+1}}
   \sum_{n=0}^Nc_n
   {\Gamma(N+(\alpha-1)n-\sigma+1)
    \over(N-n)!\,\Gamma(\alpha n-\sigma)}
   \rho^n.
\end{equation}
PMS therefore leads an equation
\begin{equation}
   \sum_{n=0}^Nc_n
   {\Gamma(N+(\alpha-1)n-\sigma+1)
    \over(N-n)!\,\Gamma(\alpha n-\sigma)}
   \rho^n=0.
\end{equation}

On the other hand, it is easy to see that FAC,
$S_N^{\rm (ODM)}-S_{N-1}^{\rm (ODM)}=0$, leads a similar equation
\begin{equation}
   \sum_{n=0}^Nc_n
   {\Gamma(N+(\alpha-1)n-\sigma)
    \over(N-n)!\,\Gamma(\alpha n-\sigma)}
   \rho^n=0.
\end{equation}

Optimization condition thus leads to an order by order deter\-mi\-na\-tion
(albeit among all pos\-sible, at most $N$ple, roots) of $\rho$ which
depends only on $N$ but inde\-pendent of $g$. On the other hand, our
convergence proof is valid if the parameter $\rho$ is appropriately
scaled with $N$, but independently of $g$. See Eq.~(\ref{mu'}) or
Eq.~(\ref{gamma'}). These facts are consistent with each other: as noticed
in Sec.~\ref{sec:opti}, optimization forces $x$ or
$\rho$ to lie within the range Eq.~(\ref{mu'}) or Eq.~(\ref{gamma'})
for which DE-ODM converges, hence selects out $\rho$ which grows
appropriately with $N$ but in a $g$ independent manner.

\section{Critical exponents in $\phi_3^4$ theory}
\label{sec:appd}
In this appendix, we apply DE-ODM to the renormalization group
functions of $\phi_3^4$ theory and compute various critical exponents.
This problem was already studied in \cite{SZ}, but our approach here
differs in the following aspect:
In \cite{SZ}, a renormalization group argument was used to relate first
the analytic structure in $g_0$ around $g_0=\infty$ ($g_0$ is the
unrenormalized coupling constant) to the behavior of the beta
function $W(g)$ near the infrared fixed point $g\simeq g^*$.
The parameter of the conformal transformation is accordingly
taken to be $\alpha=1/\omega$ ($\omega$ being the slope of the beta
function at the fixed point), in order to render $W(g(g_0(\beta)))$
analytic around $\beta=1$. The input value of $\omega$ must
somehow be guessed by e.g., other approximate methods for computing
$W(g)$. Although quite encouraging results have been found \cite{SZ},
the fact that one must assume an input value of $\omega$ (which
is one of the results being sought for) to start the calculation with,
is a somewhat disturbing feature of the approach \cite{private}.

On the other hand, as is seen in Sec.~\ref{sec:proof1},
the requirement of analyticity near $\beta=1$ is not really necessary,
if one is only interested in the convergence for finite real positive
$g$. Also, quite a wide range of conformal transformations can be used
in such a case. In this spirit one may {\it a priori\/} try a simpler
conformal transformation, directly to the renormalized coupling constant
$g$.

Here we choose $\alpha=3/2$ and $\sigma=1/2$, the same value as
in the case of the anharmonic oscillator. Assuming the renormalization
group functions to have the analyticity on the cut $g$ plane
(which is not proven to our knowledge), one concludes that DE-ODM gives a
convergent sequence for true functions.\footnote{
It is known that the perturbation series for the renormalization group
functions is asymptotic for ${\rm Re}\,g\geq0$ \cite{eck}, and the known
large order behavior of those functions \cite{bre} is consistent with the
assumed form of the imaginary part Eq.~(\ref{newdisc}) with $b=1$.}

We thus proceed as follows: take a perturbative series of, say, the
beta function
\begin{equation}
   W(g)\sim\sum_{n=0}^\infty c_ng^n,
\label{originalbeta}
\end{equation}
for which the first six non-trivial coefficients have been
calculated \cite{bak}. Then ODM with $\alpha=3/2$ and $\sigma=1/2$
gives a new series
\begin{equation}
   W_N(g,x)
   =\sum_{n=0}^Nc_ng^n{1\over x^{3n-1}}
   \sum_{k=0}^{N-n}{\Gamma(3n/2+k-1/2)\over k!\,\Gamma(3n/2-1/2)}
   \left(1-{1\over x^2}\right)^k,
\end{equation}
where $\rho$ and $\beta$ have been parameterized in terms of $x$ as in
Eq.~(\ref{conf2}). At lower orders we expect that an optimized
determination of $x_N$ gives a more accurate result than with
a particular scaling of $x_N$ with $N$. We thus employ here
the principle of minimal sensitivity, i.e., $dW_N/dx_N=0,$ for each value
of $g$, by choosing the root closest to $x_N=1$.

In Fig.~5, we plot the original beta function Eq.~(\ref{originalbeta})
up to $O(g^7)$ \cite{bak} and the function $W_7(g,x_7(g))$ obtained
in this way. A clear improvement and the appearance of an infrared fixed
point can be seen.

We applied the same method for perturbative series of various
renormalization group functions \cite{bak}. The resulting values of the
critical exponents, as a function of the order $N$, are summarized in
Table~1.\footnote{
For the definition of various critical exponents, see \cite{gui}.}
The first two columns of Table, $g^*$ and $\omega$, are the result of
the method explained above. In the last two columns, we have used
$g^*=1.43645$, the result of the seventh order resummation.
The blank means there is no solution of the optimization condition
near $x_N\sim1$.

These results can be compared with those obtained by the Borel summation
method \cite{gui}:
\bea
  &&g^*=1.416\pm0.005,\quad\omega=0.79\pm0.03,
\non\\
  &&\eta=0.031\pm0.004,\quad\gamma=1.241\pm0.0020.
\eea
Considering the simplicity of our approach (in particular, the uniform
convergence for $g\in[0,\infty]$ is not guaranteed), the agreement is
rather surprising and suggests the correctness of our assumptions.

\section{Strong coupling expansion coefficients from
standard perturbation series}
\label{sec:appe}
As shown in Sec.~\ref{sec:proof2} DE-ODM for the anharmonic oscillator
converges even in the infinite coupling constant limit. This fact can
be used to reconstruct the strong coupling expansion Eq.~(\ref{largeg})
{}from the perturbation series coefficients $c_n$ \cite{wjanke}.
Here we derive such a formula, which is more general than that given
in \cite{wjanke} and whose derivation is also somewhat simpler.

It is interesting that a quite similar formula had been proposed
earlier \cite{bonnier}, following a completely different approach.

The $N$th order DE-ODM approximant is given by \cite{GKS}
\beq
   S_N=
   \sum_{n=0}^Nc_n g^n{1\over x^{3n-1}}
   \sum_{k=0}^{N-n}{\Gamma(3n/2+k-1/2)\over k!\,\Gamma(3n/2-1/2)}
   \left(1-{1\over x^2}\right)^k.
\label{eichi}
\eeq
Substituting $x=g^{1/3}y_N$ (see Eqs.~(\ref{gamma'}), (\ref{extreme}))
into Eq.~(\ref{eichi}), we find
\begin{equation}
   S_N=
   g^{1/3}
   \sum_{n=0}^N\sum_{k=0}^{N-n}\sum_{l=0}^{k}
   c_n{\Gamma(3n/2+k-1/2)\over
       \Gamma(3n/2-1/2)(k-l)!}
   {1\over y_N^{3n+2l-1}}
   {(-1)^l\over l!}(g^{-2/3})^l.
\label{eni}
\end{equation}
A comparison with Eq.~(\ref{largeg}) then gives for the $l$th
strong coupling expansion coefficients $d_l$,
\begin{eqnarray}
   d_l&=&
  \lim_{N\to\infty}{(-1)^l\over l!}\sum_{n=0}^{N-l}\sum_{k=0}^{N-l-n}
   c_n{\Gamma(3n/2+k+l-1/2)\over k!\,\Gamma(3n/2-1/2)}
   {1\over y_N^{3n+2l-1}} \\
   &=&
   \lim_{N\to\infty}{(-1)^l\over l!}\sum_{n=0}^{N-l}
   c_n{\Gamma(N+n/2+1/2)\over
       (N-l-n)!\,\Gamma(3n/2-1/2)(3n/2+l-1/2)}
   {1\over y_N^{3n+2l-1}},  \non
\end{eqnarray}
where $y_N$ must be scaled appropriately with $N$ ($y_N =cN^\gamma$;
$1/3<\gamma<1/2$ with arbitrary positive $c$, or $y_N=cN^{1/3}$;
$c\ge c_*=0.570875\cdots$), or simply determined order by order by
optimization of $S_N$ or $d_l$.

Delta expansion for the double well potential is obtained simply by changing
$\beta\to1+1/x^2$ \cite{GKS}. From the above derivation, it is obvious that
$d_l^{\rm(DW)}=(-1)^ld_l^{\rm(AHO)}$, which is of course a correct
result, in view of Eq.~(\ref{scaling2}).

Finally, as a further check, the known strong coupling expansion
(\ref{san}) in the zero dimensional model discussed in \ref{sec:zero}
can exactly be reproduced following the same procedure.


%
\vfill\eject
\noindent
{\bf\large Table Caption}
\smallskip
\begin{description}
\item{1.} Determination of various critical exponents in $\phi_3^4$
theory by DE-ODM.
\end{description}
\bigskip
\noindent
{\bf\large Figure Captions}
\smallskip
\begin{description}
\item{1.}
The domain of analyticity $\cal D$ in $\beta$ plane,
corresponding to the cut $g$ plane. Its boundary is the curve
$\Gamma_A$ which wraps around the negative real $\beta$ axis,
and $\Gamma_B$, the upper and lower complex sections.
\item{2.}
Enlargement of the contour $\Gamma_B$ to $\Gamma'_B$,
so as to include part of the circle $\cal C$.
The crosses represent the (schematic) position of Bender-Wu singularities.
\item{3.}
The inverse image of the integration contour $\Gamma'_B$ (solid line)
and the boundary of the region Eq.~(\ref{complexg}) (dashed line) on the
$g$ Riemann surface. Straight sections of the contour image are actually
on the axis, ${\rm Arg}\,g=\pm\pi$, $\pm2\pi$, $\pm4\pi$.
\item{4.}
The contour used in the convergence proof of the zero dimensional model.
\item{5.}
Standard perturbation result for the beta function $W(g)$ (dotted line)
and DE-ODM improved one (solid line).
\end{description}
\vfill\eject
\bigskip
\begin{table}
\centering
\begin{tabular}{lllll}
\hline\hline
$N$ & $g^*$ & $\omega$ & $\eta$ & $\gamma$ \\
\hline
$2$ & ---        & ---        & ---         & $1.22847$  \\
$3$ & $1.77550$  & $0.548573$ & $0.025686$ & ---        \\
$4$ & ---        & ---        & ---         & $1.24137$  \\
$5$ & $1.48560$  & $0.699459$ & $0.030055$ & ---        \\
$6$ & ---        & ---        & ---         & $1.24366$  \\
$7$ & $1.43645$  & $0.751379$ &             &            \\
\hline\hline
\end{tabular}
\bigskip
\caption{Determination of various critical exponents in $\phi_3^4$
theory by DE-ODM.}
\end{table}
\vfill
\end{document}